\documentclass[a4paper,fleqn,usenatbib]{mnras} 
\usepackage[T1]{fontenc}
\usepackage{ae,aecompl}
\usepackage{mathtools,tabu}
\usepackage{amsmath,amssymb}
\usepackage{makecell} 
\usepackage{upgreek} 
\usepackage{float} 

\newcommand{\avg}[1]{\left\langle{#1}\right\rangle}
\newcommand{\comment}[1]{}
\newcommand{\beq}{\begin{equation}}
\newcommand{\eeq}{\end{equation}}
\newcommand{\beqa}{\begin{equation}\begin{aligned}}
\newcommand{\eeqa}{\end{aligned}\end{equation}}
\newcommand{\bit}{\begin{itemize}}
\newcommand{\eit}{\end{itemize}}

\newcommand{\hiGpc}{h^{-1} \rm Gpc}
\newcommand{\hiMpc}{h^{-1} \rm Mpc}
\newcommand{\hikpc}{h^{-1} \rm kpc}
\newcommand{\hiMsun}{h^{-1} \rm M_\odot}
\newcommand{\Msun}{\rm M_\odot}

\newcommand{\Cov}{{\rm Cov}}
\newcommand{\DS}{\Delta\Sigma}
\newcommand{\Sigcrit}{\Sigma_{\rm crit}}

\newcommand{\clhh}{C_\ell^{\rm hh}}
\newcommand{\clkk}{C_\ell^{\kappa\kappa}}
\newcommand{\clhk}{C_\ell^{\rm h\kappa}}

\newcommand{\clSS}{C_\ell^{\Sigma\Sigma}}
\newcommand{\clhS}{C_\ell^{\rm h\Sigma}}

\newcommand{\fsky}{f_{\rm sky}}
\newcommand{\nside}{n_{\rm side}}
\newcommand{\psrc}{p_{\rm src}}
\newcommand{\hJ}{\hat{J}}
\newcommand{\thetamin}{\theta_{\rm min}}
\newcommand{\thetamax}{\theta_{\rm max}}

\newcommand{\tu}{\tilde{u}}

\newcommand{\rp}{r_{\rm p}}

\newcommand{\Pmm}{P_{\rm mm}}
\newcommand{\Phm}{P_{\rm hm}}

\DeclareRobustCommand{\rchi}{{\mathpalette\irchi\relax}}
\newcommand{\irchi}[2]{\raisebox{\depth}{$#1\chi$}} 

\newcommand{\chis}{\rchi_{\rm s}}
\newcommand{\chih}{\rchi_{\rm h}}
\newcommand{\chiLSS}{\rchi_{\rm lss}}

\newcommand{\chimin}{\rchi_{\rm min}}
\newcommand{\chimax}{\rchi_{\rm max}}

\newcommand{\zs}{z_{\rm s}}
\newcommand{\zh}{z_{\rm h}}
\newcommand{\zLSS}{z_{\rm lss}}

\newcommand{\Ci}{{\rm Ci}}
\newcommand{\Si}{{\rm Si}}

\newcommand{\nh}{n_{\rm h}^{\rm (2D)}}
\newcommand{\ns}{n_{\rm s}^{\rm (2D)}}

\newcommand{\dd}{{\rm d}}
\newcommand{\xihm}{\xi_{\rm hm}}
\newcommand{\sigmaf}{\sigma_{f}}

\title[Cluster lensing covariance matrices]
{Covariance matrices for galaxy cluster weak lensing: from virial regime to uncorrelated large-scale structure}
\author[Wu et al.]{Hao-Yi Wu,$^{1}$\thanks{Email: wu.3863@osu.edu} 
David H. Weinberg,$^{1}$
Andr\'{e}s N. Salcedo,$^{1}$
Benjamin D. Wibking,$^{1}$
\newauthor 
and Ying Zu$^{2}$\\
$^{1}$  Department of Physics, Department of Astronomy, and Center for Cosmology and Astro-Particle Physics, \\ \ \ The Ohio State University, Columbus, OH 43210, USA\\
$^{3}$ Department of Astronomy, School of Physics and Astronomy, Shanghai Jiao Tong University, Shanghai 200240, China
}
\pubyear{2019}
\begin{document}

\label{firstpage}
\pagerange{\pageref{firstpage}--\pageref{lastpage}}
\maketitle

\begin{abstract}
Next-generation optical imaging surveys will revolutionise the observations of weak gravitational lensing by galaxy clusters and provide stringent constraints on growth of structure and cosmic acceleration.  In these experiments, accurate modelling of covariance matrices of cluster weak lensing plays the key role in obtaining robust measurements of the mean mass of clusters and cosmological parameters.   We use a combination of analytical calculations and high-resolution N-body simulations to derive accurate covariance matrices that span from the virial regime to linear scales of the cluster-matter cross-correlation. We validate this calculation using a public ray-tracing lensing simulation and provide a software package for calculating covariance matrices for a wide range of cluster and source sample choices.  We discuss the relative importance of shape noise and density fluctuations, the impact of radial bin size, and the impact of off-diagonal elements.  For a weak lensing source density $n_{\rm s}=10$ arcmin$^{-2}$, shape noise typically dominates the variance on comoving scales $\rp \la 5\ \hiMpc$.  However, for $n_{\rm s}=60$ arcmin$^{-2}$, potentially achievable with future weak lensing experiments, density fluctuations typically dominate the variance at $\rp \ga 1\ \hiMpc$ and remain comparable to shape noise on smaller scales.  
\end{abstract}


\section{Introduction}

Understanding the origin of cosmic acceleration requires measurements of both the expansion rate of the Universe and the growth rate of large-scale structure \citep[LSS, see e.g.][for reviews]{Frieman08,Weinberg13,Huterer15}.
The number counts of galaxy clusters as a function of mass and redshift are sensitive to both expansion rate and the growth of structure \citep[e.g.][]{Holder01,Haiman01,Miller01,Vikhlinin09,Rozo10,Mantz10,Benson13,Planck13Cluster,Planck15Cluster,Mantz14}; also see \cite{Allen11} for a review.   
With optical surveys, one can identify clusters to lower mass thresholds at high redshifts relative to X-ray or submillimeter surveys.  Regardless of identification method, calibrating the mass scale associated with observed cluster properties plays a key role in extracting cosmological information from galaxy clusters.  
In optical imaging surveys, one can use weak gravitational lensing to constrain mean cluster mass density profiles
\citep[e.g.][]{Hoekstra07, Mahdavi08, Zhang08, Vikhlinin09,Zhang10, Rozo11wl, Oguri11, BeckerKravtsov11, Applegate14, vdlinden14b, Hoekstra15, Kohlinger15, OkabeSmith16, Melchior17, Simet17, Medezinski18, Murata18, Dietrich19, Miyatake19}.
These profiles can often be measured with high precision far beyond the cluster virial radius, to scales of tens of Mpc, where they probe the linear regime of the cluster-matter cross-correlation function \citep{Sheldon09,Zu14,Melchior17,McClintock19}.  The combination of cluster space densities and weak lensing profiles is in many ways analogous to the combination of galaxy clustering and galaxy-galaxy lensing, but applied to the high mass end of the dark matter halo population.  In these experiments, accurate modelling of the signals and the covariance matrices is essential for obtaining robust constraints on cluster masses and cosmological parameters.

To date, most of the weak gravitational lensing measurements of galaxy clusters are dominated by shape noise; that is, because of the modest number density of source galaxies, the noise of lensing is dominated by the intrinsic ellipticities of source galaxies.  However, for the next-generation optical surveys like the Large Synoptic Survey Telescope (LSST, \citealt{LSST}), {\rm Euclid} \citep{Euclid}, and the {\rm Wide Field Infrared Survey Telescope} ({\rm WFIRST}, \citealt{WFIRST}), the number density of source galaxies will be significantly higher and shape noise will no longer dominate.  Instead, the density fluctuations related to the intrinsic variation of halo density profiles will dominate the uncertainties at small scales, and the contribution of uncorrelated LSS will dominate the uncertainties at large scales.  In this paper, we detail how to calculate covariance matrices for cluster lensing at all scales with and without shape noise, and we provide a software package that can calculate covariance matrices for a wide range of survey assumptions.\footnote{\href{https://github.com/hywu/cluster-lensing-cov}{https://github.com/hywu/cluster-lensing-cov}}

The covariance matrices of cluster lensing or galaxy-galaxy lensing can be calculated directly from data using 
jackknife methods \citep[e.g.][]{McClintock19}, 
analytical formulae \citep[e.g.][]{Hoekstra01,Hoekstra03,Hoekstra11,Marian15,Singh17}, 
or simulations \citep[e.g.][]{Shirasaki17,Harnois18}.  
In this paper, we focus on the latter two approaches.  The literature for cluster lensing covariance matrices is relatively small.  \cite{Gruen15} quantified the impact of intrinsic variation of halo density profiles on the covariance matrices of cluster weak lensing by combining analytical calculations with numerical simulations.  However, as we will explain in Section~\ref{sec:pitfalls}, they quantified the halo-to-halo covariance rather than the patch-to-patch covariance, and the latter is relevant for stacked cluster lensing.

The covariance matrices of galaxy-galaxy lensing are more widely studied.  Galaxy-galaxy lensing is analogous to cluster lensing and refers to using  galaxies (lower-mass haloes) instead of galaxy clusters as lenses. 
\cite{Singh17} provided analytical formulae for galaxy-galaxy lensing covariance matrices. Using observed and simulated catalogues, they compared various sources of errors and demonstrated the importance of subtracting lensing signals around random points. Their analysis at small scales is dominated by shape noise
and is therefore not directly applicable for the next-generation cluster lensing surveys, where the non-Gaussian covariance will be significant compared with shape noise.
\cite{Shirasaki17} used simulated full-sky weak lensing maps to compare various methods for calculating covariance matrices and demonstrated the accuracy of jackknife covariance.
\cite{Harnois18} developed large-volume simulations for calculating the cross-probe covariance for cosmic shear, galaxy-galaxy lensing, and galaxy clustering.  
Other earlier work includes \cite{Jeong09,Harnois12,Marian15}.
These papers focus on lower-mass haloes and large scales; therefore, they do not provide the small-scale covariance matrices we need for cluster lensing mass calibration.

Because of the reasons stated above, for the next-generation cluster lensing measurements by LSST, {\rm Euclid}, and {\rm WFIRST}, we need to come up with a new treatment for covariance matrices.  In this paper, we first compare the analytical covariance assuming Gaussian random fields with the covariance from simulated lensing maps.  We then identify the scales where the Gaussian-field covariance is insufficient, and we add corrections based on high-resolution N-body simulations. We provide readers with a set of user-friendly equations with short and heuristic derivations, a software package, and tabulated results from simulations.

Readers might ask why we do not use simulated gravitational lensing ray-tracing maps to calculate covariance matrices at all scales.  For example, \cite{Takahashi17} and \cite{Shirasaki17} produced full-sky lensing maps and used them to study the covariance matrices of galaxy-galaxy lensing, and we use these maps to cross-check our calculations.  However, because of the substantial computational resources required, these maps are limited to one cosmology and are unable to resolve the inner profiles of clusters. We need the contribution from very small scales (to account for the intrinsic variation of density profiles) to very large scales (to account for the LSS contribution).  For the former, we use high-resolution N-body simulation boxes to characterise the non-linear evolution of dark matter haloes.  For the latter, it is impractical to use N-body simulation boxes because we would need to include dark matter particles extending to a few Gpc, while analytical calculations are much more efficient in this regime.  We will show that this approach agrees with the ray-tracing simulations of \cite{Takahashi17} at intermediate and large scales.

We use two sets of public simulations: the ray-tracing lensing maps from \cite{Takahashi17} and the N-body simulations from Abacus Cosmos \citep{Garrison18}. Table~\ref{tab:sims} summarises the specifications of the two simulations.  The difference in their cosmological parameters leads to negligible difference in covariance matrices.

Throughout this work, we use $M_{\rm 200m}$, the mass defined by the radius within which the overdensity is 200 times the mean density of the Universe.  All distances are in {\em comoving} $\hiMpc$; $\chi$ denotes the line-of-sight distances, and $\rp$ denotes projected (transverse) distances.  Since cluster lensing covariance matrices are inversely proportional to the survey area, we normalise all of the covariance values to those of a volume of 1 $(\hiGpc)^3$. Table~\ref{tab:notations} summarises the notations used in this work.  We use ``variance'' to indicate the diagonal elements of a covariance matrix.

This paper is organised as follows.  
Section~\ref{sec:basics} provides the basic equations of cluster weak lensing.
Section~\ref{sec:lensing_maps} describes how we measure covariance matrices from lensing simulations and possible pitfalls in this procedure. 
In Section~\ref{sec:cov_gammat} we focus on large-scale covariance matrices for tangential shear assuming Gaussian random fields, and in Section~\ref{sec:cov_DS} we present the analogous calculations for the excess surface density.
In Section~\ref{sec:grafting} we use high-resolution N-body simulations to calculate small-scale covariance matrices and combine simulations with analytical calculations. 
We discuss the correlation between radial bins in Section~\ref{sec:off-diagonal}.
In Section~\ref{sec:discussion} we discuss shape noise, the dependence of covariance matrices on halo mass and redshift, and cross-mass covariance matrices.  Section~\ref{sec:summary} summarises our results.

\section{Background of cluster weak lensing}\label{sec:basics}

Below we briefly describe the basic equations for cluster weak lensing. We refer interested readers to e.g.~\cite{BartelmannSchneider01,Weinberg13,Kilbinger15}  for comprehensive reviews.

The gravitational lensing signal is quantified by the Jacobian matrix for the coordinate transformation from the source plane to the lens plane,
\beq
\begin{pmatrix}
  1-\kappa-\gamma_1 & -\gamma_2 \\
  -\gamma_2 & 1-\kappa+\gamma_1
 \end{pmatrix}  \ ,
\eeq
where $\kappa$ corresponds to convergence, and ($\gamma_1$, $\gamma_2$) corresponds to the two components of shear.  We ignore higher-order effects in this work.  If we assume that the line-of-sight dimension of a dark matter halo is much smaller than the distance between the observer and the source galaxy (the thin-lens approximation), the azimuthally averaged convergence is related to the surface density ($\Sigma$) of a halo via \beq
\kappa(\rp) = \frac{\Sigma(\rp)}{\Sigcrit}  \ , 
\eeq
where $\rp$ is the projected distance on the lens plane in comoving units, and $\Sigcrit$ is the critical surface density defined as 
\beq
\Sigcrit(\zs, \zh) = \frac{c^2}{4\uppi G}\frac{\chis}{\chih (\chis-\chih)(1+\zh)} \ ,
\eeq
where $\chi$ denotes comoving distances; we use subscripts h and s to denote the redshifts or distances of haloes (galaxy clusters) and sources (background galaxies).  Note that we have $(1+\zh)$ in the denominator because we use comoving units; if one uses physical units, $\Sigcrit$ would differ by $(1+\zh)^2$ and would usually be written in terms of angular diameter distances.

The two shear components depend on the choice of coordinate system, and a physical quantity is the tangential shear,
\beq
\gamma_t = - \gamma_1 \cos(2\phi) - \gamma_2 \sin(2\phi)  \ ,
\eeq
where $\phi$ is the position angle of the source galaxy with respect to the cluster centre.  The azimuthally averaged tangential shear is related to the excess surface density ($\DS$) via
\beq
\gamma_t = \frac{\DS}{\Sigcrit} = \frac{\Sigma(<\rp) - \Sigma(\rp)}{\Sigcrit} = \kappa(<\rp) - \kappa(\rp)  \ .
\label{eq:gammat_kappa}
\eeq
The observable is the reduced shear of individual source galaxies, $g_i = \gamma_i/(1-\kappa)$, $i=1,2$. 
In the weak lensing regime, $\kappa\ll 1$, and $\gamma_t$ is the observable for practical purposes.  
In the context of cluster lensing, one often converts $\gamma_t$ to $\DS$ because the latter can be interpreted physically as the excess surface density profiles of clusters.

The two quantities, $\DS$ and $\gamma_t$, have different advantages and disadvantages.
For a given lens, $\DS$ is independent of source redshift, while $\gamma_t$ is higher for a higher source redshift.  On the other hand, to calculate $\DS$ from the observed $\gamma_t$ one needs to assume a cosmology and know the redshifts of sources and lenses.  In this work, we will first focus on the covariance matrices of $\gamma_t$ in Section~\ref{sec:cov_gammat} because the contribution from LSS to the noise of $\gamma_t$ is easier to understand. We will then discuss the analogous covariance matrices of $\DS$ in Section~\ref{sec:cov_DS}.

For an ensemble of galaxy clusters at a given redshift, the mean $\DS(\rp)$ profile can be computed from the halo-matter correlation function $\xihm$ using
\beqa
\DS(\rp) =& \bar{\rho}\bigg[\frac{4}{\rp^2}\int_0^{\rp} {\rp^\prime}\dd{\rp^\prime} \int_0^\infty  
\dd\chi\ \xihm\left(\sqrt{{\rp^\prime}^2+\chi^2}\right)
\\ 
& - 2 \int_0^\infty \dd\chi\ 
\xihm\left(\sqrt{\rp^2+\chi^2}\right)
\bigg]  \ .
\label{eq:DS_realspace}
\eeqa
This real-space expression is useful for calculating mean profiles of $\DS$ or $\gamma_t$, while the equivalent Fourier-space expression (equation~\ref{eq:gammat_hankel} below) is more useful for calculating covariance matrices.\footnote{In simulations, if we calculate $\xihm$ in a volume and use this formula to calculate the corresponding $\DS$, we tend to underestimate the covariance between different simulation volumes, because $\xihm$ averages over 3D and does not include the variance due to different lines of sight (see \citealt{Salcedo19}).}

To calculate $\xihm$, we use the approach in \cite{HayashiWhite08}.  On scales smaller than the virial radius, $\xihm$ represents the average density profile. We use an Navarro-Frenk-White (NFW) profile \citep{NFW97}, $\rho_{\rm NFW}(r)$, with the concentration--mass relation from \cite{Bhattacharya13}.  We assume no scatter between concentration and mass.  On scales much greater than the virial radius, $\xihm$ is the linear matter correlation function $\xi_{\rm lin}$ multiplied by the halo bias factor $b$.  For intermediate scales, we use the larger of the two; that is,
\beq
\xihm = {\rm max}(\xi^{\rm 1h}, \xi^{\rm 2h}) \ ,
\eeq
where
\beqa
\xi^{\rm 1h} &= \frac{\rho_{\rm NFW}(r)}{\bar{\rho}} - 1 \\
\xi^{\rm 2h} &= b(M) \xi_{\rm lin}(r) \ .
\eeqa

\begin{table*}
\centering
\caption{Simulations used in this work. See Sections~\ref{sec:takahashi} and \ref{sec:abacus} for details for Takahashi and Abacus, respectively.}
\setlength{\tabcolsep}{3mm}
\begin{tabular}{ccccc}
\hline
\rule[-2mm]{0mm}{6mm} Simulation & Type   & Cosmology &  Our usage 
 &  Subsamples used in this paper
\\\hline\hline
\rule[-2mm]{0mm}{6mm} Takahashi
    & \makecell[cl]{full-sky lensing maps  \\ and haloes in lightcones} 
    & \makecell[cl]{$\Omega_m=0.279$\\$\sigma_8=0.82$\\$h=0.7$}
    & \makecell[cl]{validating calculations at\\ large and intermediate scales}
    & \makecell[cl]{lightcones with $\fsky$=1/48 (680 deg$^2$),\\ 480 realisations at $z\approx$ 0.1, 0.3, 0.5, 0.7}
    \\\hline
\rule[-2mm]{0mm}{6mm}  Abacus   
    & \makecell[cl]{particles and haloes\\ in boxes} 
    & \makecell[cl]{$\Omega_m=0.314$\\$\sigma_8=0.83$\\$h=0.67$}
    & \makecell[cl]{calculating covariances at \\ small scales}
    & \makecell[cl]{subvolumes of $720^3/9 \ (\hiMpc)^3$,\\ 180 realisations at $z=$ 0.3, 0.5, 0.7}
    \\\hline
\end{tabular}
\label{tab:sims}
\end{table*}

\begin{table*}
\centering
\caption{Meanings of the symbols used in this work.}
\setlength{\tabcolsep}{3mm}
\begin{tabular}{lll}
\hline
\rule[-2mm]{0mm}{6mm} Symbol & Meaning & Unit \\\hline\hline
\rule[-2mm]{0mm}{6mm} $\chih$, $\chis$, $\chiLSS$ & comoving distances between the observer and the halo, the source, or LSS & $\hiMpc$ \\
\rule[-2mm]{0mm}{6mm} $\theta$& angular separation between the halo centre and the source galaxy & radian \\
\rule[-2mm]{0mm}{6mm} $\rp$& projected comoving distance on the lens plane, $\rp=\chih\theta$ & $\hiMpc$ \\
\rule[-2mm]{0mm}{6mm} $\nh$ & surface number density of haloes (galaxy clusters) & sr$^{-1}$ \\
\rule[-2mm]{0mm}{6mm} $\ns$ & surface density of sources (background galaxies) & sr$^{-1}$ \\
\rule[-2mm]{0mm}{6mm} $\psrc(\chis)$ & redshift distribution of sources, $\int_0^\infty \dd\chis \psrc(\chis)=1$ & $h\ {\rm Mpc^{-1}}$ \\
\rule[-2mm]{0mm}{6mm} $\bar{\rho}$ & comoving mean density of the Universe & $h^2\rm \Msun Mpc^{-3}$ \\
\rule[-2mm]{0mm}{6mm} $\sigma_\gamma$ & noise for shear & dimensionless \\
\hline
\end{tabular}
\label{tab:notations}
\end{table*}

\section{Covariance matrices from weak lensing maps}\label{sec:lensing_maps}

We start by measuring the covariance matrices from the weak lensing maps of \cite{Takahashi17} and discussing pitfalls associated with these measurements.  We will later use these measurements to cross-check our calculations combining analytical formulae and N-body simulations.

\subsection{Takahashi ray-tracing simulations}\label{sec:takahashi}

We use the publicly available lensing maps produced by 
\citet[][also see \citealt{Shirasaki15,Shirasaki17}]{Takahashi17}\footnote{\href{http://cosmo.phys.hirosaki-u.ac.jp/takahasi/allsky_raytracing/}{http://cosmo.phys.hirosaki-u.ac.jp/takahasi/allsky\_raytracing/}}.
These authors built full-sky lightcones using N-body simulations based on {\sc Gadget2} \citep{Springel05}.
The N-body simulations are based on a flat $\Lambda$CDM cosmology consistent with WMAP9  \citep{Hinshaw13},
$\Omega_m$ = 0.279;
$\Omega_{\rm DE}$ = 0.721;
$\Omega_b$ = 0.046;
$h$ = 0.7;
$\sigma_8$ = 0.82;
$n_s$ = 0.97.
The dark matter haloes are identified using {\sc Rockstar} \citep{Behroozi13rs}.  For the lightcones, starting from $z=0$, for every 450 $\hiMpc$ the authors used a different N-body simulation volume with progressively lower resolution.  For the lensing simulations, they set a source plane every 150 $\hiMpc$ and generated full-sky maps of convergence, shear, and rotation using ray-tracing calculations for all the dark matter in front of the source plane.  They provided maps in the Healpix format with resolution $\nside$ = 4096, 8192, and 16384.  In our calculations, we use the maps with $\nside$ = 4096 (corresponding to $2\times10^8$ pixels in the full sky), and the angular resolution of each pixel is $2.5\times10^{-4}$ radian or 0.86 arcmin.  We have checked that the $\nside$=8192 maps give nearly identical results.

Since the lightcone exhibits discontinuities every 450 $\hiMpc$, 
we choose lens redshift bins that avoid these discontinuities; that is, a halo sample we choose comes from one original N-body simulation box and thus has continuous LSS.
For our fiducial calculation, we use haloes from lens plane 10, which corresponds to a comoving distance range (1350, 1500) $\hiMpc$ and a redshift range (0.508, 0.574). The haloes in this redshift range are generated from an N-body simulation of box size 1800 $\hiMpc$ with 2048$^3$ particles; the mass resolution is $5.3\times10^{10}\ \hiMsun$, and the softening length is 32 $\hikpc$.  We use source plane 18, which corresponds to a redshift range (1.218, 1.318); see Tables 1 and 2 in \cite{Takahashi17}.

We divide the full-sky catalogues into 48 equal-area samples.  We use 10 of the 108 realisations provided by the authors; that is, we have 480 realisations of a survey with $\fsky=1/48$ (860 deg$^2$).  To calculate the averaged $\gamma_t$ profiles, we take all the haloes with $M_{\rm 200m}\ge 10^{14}\ \hiMsun$ in each sample (on average 785 haloes per sample) and cross-correlate them with all pixels in the shear map using the publicly available code {\sc TreeCorr}\footnote{\href{https://github.com/rmjarvis/TreeCorr}{https://github.com/rmjarvis/TreeCorr}}.
In addition, we generate random points in the same area (20 times the number of haloes) and calculate the $\gamma_t$ profile.   We subtract the shear signal around random points from the cluster lensing shear signal.  In this way, we obtain the mean shear profiles of $\gamma_t$ for each patch of $\fsky=1/48$, and we calculate the covariance matrix of these 480 realisations.  We use 5 logarithmically-spaced bins per decade of angular separation (15 bins for  $5\times10^{-4} \le \theta \le 0.5$).    As the number of realisations (480) is much greater than the number of bins (15), we do not need to correct the inverse  covariance as in \cite{Hartlap07}.

\subsection{Pitfalls of calculating covariance matrices}\label{sec:pitfalls}

\begin{figure}
\centering
\includegraphics[width=0.5\textwidth]{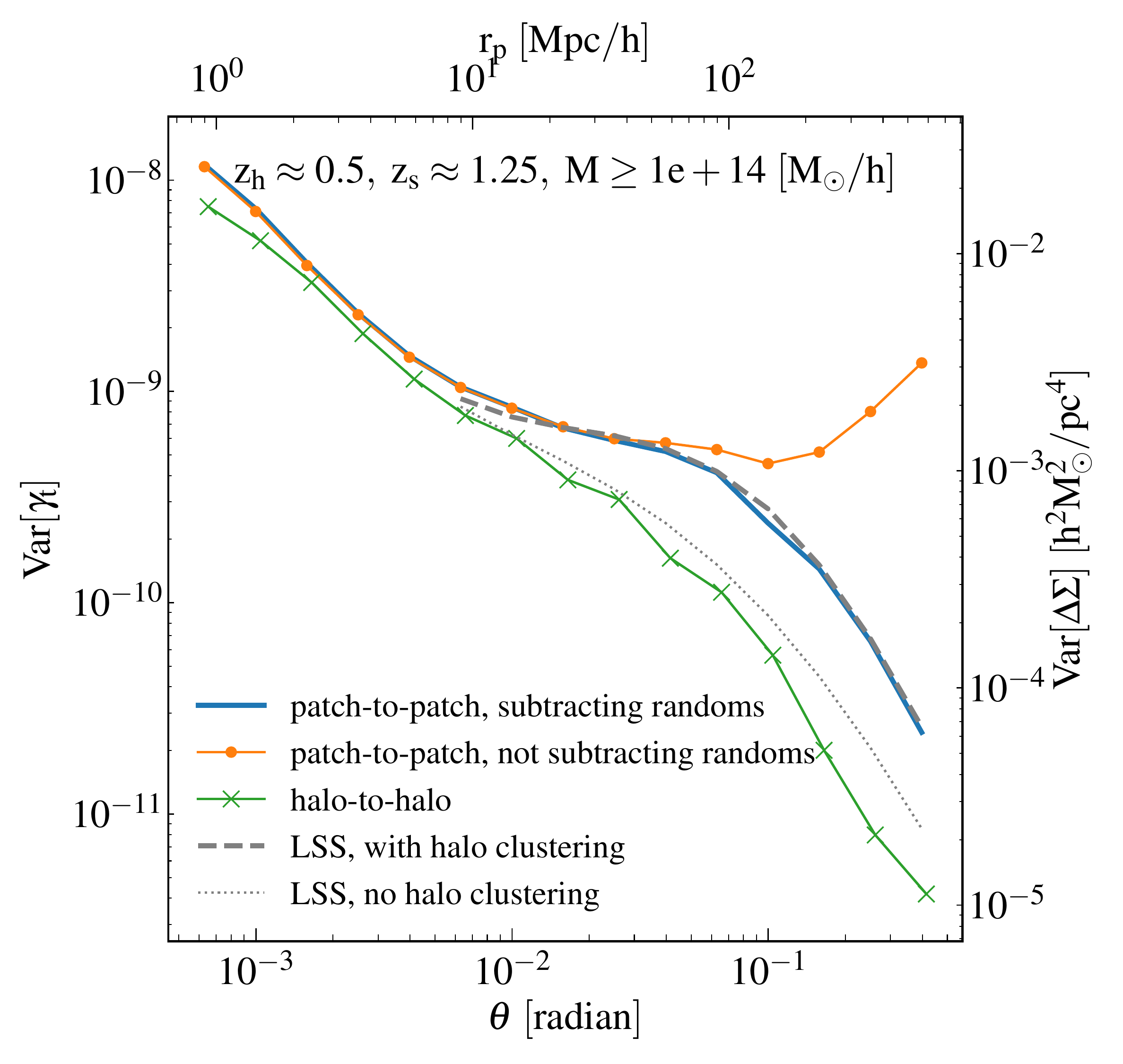}
\caption{Pitfalls of calculating cluster lensing covariance matrices, demonstrated using the Takahashi lensing maps with no shape noise.  The blue curve shows the correct method: calculating covariance matrices using multiple realisations of the same survey area; for each realisation, the lensing signal around random points is subtracted.  Not subtracting the lensing signal around random points would lead to erroneous variance at large scales (orange).   Using the halo-to-halo variance instead of the patch-to-patch variance would lead to underestimation of the large-scale variance (green). The grey dash curve corresponds to the analytical variance due to LSS and halo clustering (consistent with the patch-to-patch variance at large scales), while the grey dotted curve corresponds to the variance due to LSS without halo clustering (consistent with the halo-to-halo variance at large scales). Small-scale behaviours will be explored in the rest of this paper.}
\label{fig:var_pitfalls}
\end{figure}

Before going into the details of covariance matrix calculations, we discuss two possible pitfalls of calculating and interpreting cluster lensing covariance matrices from simulations.  Fig.~\ref{fig:var_pitfalls} shows an example for calculating the covariance matrix from the Takahashi simulations (for the fiducial calculation described in Section~\ref{sec:takahashi}).

The first pitfall is calculating covariance matrices without subtracting the lensing signal around random locations.  \citet{Mandelbaum13} demonstrated that when measuring the shear signal around galaxies, subtracting the shear signal around random lenses can remove shear systematics \citep[also see][]{Mandelbaum05,Mandelbaum06}.  We find that this subtraction is not only necessary for observations but also necessary for simulated shear maps.  Fig.~\ref{fig:var_pitfalls} shows that, even when using simulations free of systematics, the patch-to-patch variance at large scales is erroneously high if one does not subtract the mean random signal (orange).  A similar excess at large scales has also been shown in \cite{Singh17} using real sources from the Sloan Digital Sky Survey.  Subtracting the shear at random points removes the impact of over- or under-densities on the scales of the survey and is roughly analogous to using a \cite{LandySzalay93} estimator for the correlation function.  For small scales, the shear signal centred on random points is small compared to that centred on clusters, and this subtraction makes negligible difference.

The second pitfall is using the halo-to-halo covariance instead of the patch-to-patch covariance.  One obvious way to estimate the covariance matrix for a sample of clusters, either observed or simulated, is to measure the $\gamma_t$ or $\DS$ profile cluster by cluster, compute the covariance matrix of these profiles, and divide by the cluster number \citep[e.g.,][]{Gruen15}.  However, this halo-to-halo covariance generally underestimates the true uncertainty of the mean profile of a stacked lens sample.  Instead, the covariance matrix should be computed from the mean profiles derived from patches of sky that are substantially larger than the largest scales being measured.  Fig.~\ref{fig:var_pitfalls} demonstrates this point using the Takahashi simulations: at scales $\theta \ga 10^{-2}$ radian ($\rp \ga 10\ \hiMpc$), the variance of $\gamma_t$ from halo to halo divided by the number of haloes in a patch (green) is smaller than the variance among the multiple patches of 860 deg$^2$ (blue).  It is this latter variance that provides an estimate of the uncertainty on the mean $\gamma_t$ profile for an 860 deg$^2$ survey.

At small scales, the halo-to-halo and the patch-to-patch estimates of the covariance are similar. The difference occurs at large scales, where the dominant contribution to the covariance is from foreground and background LSS that is not correlated with clusters themselves.  The grey dash curve in Fig.~\ref{fig:var_pitfalls} shows the analytical prediction for this LSS contribution (equation~\ref{eq:cov_lss} below), which agrees with the patch-to-patch simulation results.  The grey dotted curve corresponds to equation~(\ref{eq:cov_lss}) without the halo clustering term $\clhh$ and is close to the halo-to-halo variance.  Physically, the excess variance due to $\clhh$ arises because clustered haloes do not independently sample the foreground and background cosmic shear signal.   Mathematically, at large scale, in addition to halo shot noise, the halo-to-halo covariance includes the three-point function $\avg{\delta_h \gamma_t \gamma_t'}$, which is negligible, while the patch-to-patch covariance includes the four-point function $\avg{\delta_h \delta_h' \gamma_t \gamma_t'}$, which is non-negligible and describes the effect of halo clustering.  Therefore, the halo clustering affects the latter but not the former.

For a sample of individual massive clusters widely separated in the sky, the halo-to-halo and the patch-to-patch variances are equivalent by definition.  In this case, halo clustering is much smaller than shot noise. Thus, the effect illustrated in Fig.~\ref{fig:var_pitfalls} is not important for individual cluster measurements like Weighing the Giants \citep{vdLinden14}.  However, it will be relevant for surveys that measure stacked weak lensing profiles to large scales for large cluster samples over contiguous areas of the sky, such as DES, LSST, {\rm Euclid}, and {\rm WFIRST}.

\vspace{-5mm}
\section{Analytical Gaussian-field covariance for tangential shear ($\lowercase{\gamma_t}$)}\label{sec:cov_gammat}

We start by calculating the covariance of $\gamma_t$ derived from linear theory, with the assumption that the underlying matter density and halo number density follow Gaussian random fields.  Such an assumption is only valid at large scales.  We call such covariance ``Gaussian-field'' covariance.  
In this section we focus on the covariance of $\gamma_t$, and in Section~\ref{sec:cov_DS} we will discuss the covariance of $\DS$.
The Gaussian-field covariance of $\gamma_t$ is easier to derive and interpret because the contribution from LSS can be intuitively understood as an additional noise of $\gamma_t$.

To ensure readability, we write equations in their simplest forms, i.e., assuming no weighting for lenses or sources (see e.g.~\citealt{Shirasaki18} for equations incorporating weighting).  Our notations are similar to those in \cite{Jeong09} and \citet{Marian15}, and we use the angular power spectrum $C_\ell$ to describe the spatial distributions of lenses and sources.  Some authors \citep[e.g.][]{Singh17} use the two-dimensional power spectrum, $P(k_\perp)$, which is equivalent to $C_\ell$ on scales where sky curvature is negligible ($\ell \gg 1$).  While $C_\ell$ arises from the spherical harmonics transform of a field on a sphere and is usually dimensionless, $P(k_\perp)$ describes the perpendicular (transverse) modes in a 3D field assuming a flat geometry and usually has the dimension of distance squared.

\subsection{Covariance matrices from Fourier space}

Weak lensing covariance matrices are more easily derived in Fourier space than in real space because in Fourier space, $\gamma_t$ and $\kappa$ only differ by a phase and thus have the same power spectrum for large $\ell$ \citep[see e.g.][]{Kilbinger15, Kilbinger17}.  In contrast, in real space, they are related via equation~(\ref{eq:gammat_kappa}).

We can analytically calculate the mean tangential shear of clusters using a Hankel transform of the halo-lensing power spectrum:
\beq
\avg{\gamma_t}(\theta) = \int\frac{\ell \dd\ell}{2\uppi} \clhk J_2(\ell \theta) \ ,
\label{eq:gammat_hankel}
\eeq
where $\clhk$ is the cross angular power spectrum between haloes and shear (Section~\ref{sec:cl}), 
and $J_2$ is the Bessel function of order 2.  
This $J_2$ arises from the fact that $(\gamma_1, \gamma_2)$ is a spin-2 field; for $\avg{\kappa}$ (scalar), we simply replace $J_2$ by $J_0$.  Equation~(\ref{eq:gammat_hankel}) is equivalent to equation~(\ref{eq:DS_realspace}) based on the real-space correlation function $\xihm$. 
While the Fourier-space expression is more useful for understanding the covariance matrix derivation, the real-space expression is easier to use for calculating the mean lensing signal.

If we assume that both the halo number overdensity and the matter overdensity follow Gaussian random fields, then the covariance of $\gamma_t$ is given by (see e.g.~\citealt{Jeong09} and Appendix~\ref{app:cov})
\beqa
\Cov^{\rm Gauss}\bigg[ \gamma_t(\theta_1), \gamma_t(\theta_2) \bigg] 
= \frac{1}{4\uppi\fsky} \int \frac{\ell \dd\ell}{2\uppi}
\hJ_2(\ell \theta_1)\hJ_2(\ell \theta_2)\times \\ 
\left[\left(\clhh+\frac{1}{\nh}\right)\left(\clkk+\frac{\sigma_\gamma^2}{\ns}\right) +\left(\clhk\right)^2 \right] 
\ . 
\label{eq:cov_gammat}
\eeqa
Here $\fsky$ is the sky fraction of the survey; 
$\nh$ and $\ns$ are the surface number densities of haloes and sources in the unit of sr$^{-1}$;
$\clhh$, $\clkk$, $\clhk$ are the angular power spectra of halo-halo, lensing-lensing, and halo-lensing, respectively (see Section~\ref{sec:cl}).
We use the bin-averaged $J_2$ to take into account the finite radial bin size (e.g.~equation 26 in \citealt{Jeong09}):
\beq
\hat J_2(\ell,\thetamin, \thetamax) = \frac{1}{\uppi(\thetamax^2 - \thetamin^2)} \int_{\thetamin}^{\thetamax} J_2(\ell \theta) 2\uppi \theta \dd\theta   \ . 
\eeq 
Appendix~\ref{app:bessel} presents the properties of the bin-averaged Bessel function and the impact of bin size.  We give the analytical expressions for $\clhh$, $\clkk$, $\clhk$ in terms of 3D power spectra using the Limber approximation in Section~\ref{sec:cl}, and we provide the derivations in Appendix~\ref{app:limber}.  
The halo model we use is described in Appendix~\ref{app:halo_model}.
Appendix~\ref{app:demo_shot_noise_vs_clustering} demonstrates the relative importance of $\clhh$, $1/\nh$, $\clkk$, and $\sigma_\gamma^2/\ns$.

We use equation~(\ref{eq:cov_gammat}) for modelling the Gaussian-field variance.  This equation first appeared in \cite{Jeong09} and was used to calculate the galaxy-galaxy lensing covariance matrix at several hundred Mpc scales and to assess the sensitivity of galaxy-galaxy lensing to primordial non-Gaussianity.  In this work, we use it to calculate cluster lensing at the scales of tens to 100 Mpc.


\subsection{Interpreting the three components of the covariance matrix}

\begin{figure*}
\centering
\includegraphics[width=1.0\textwidth]{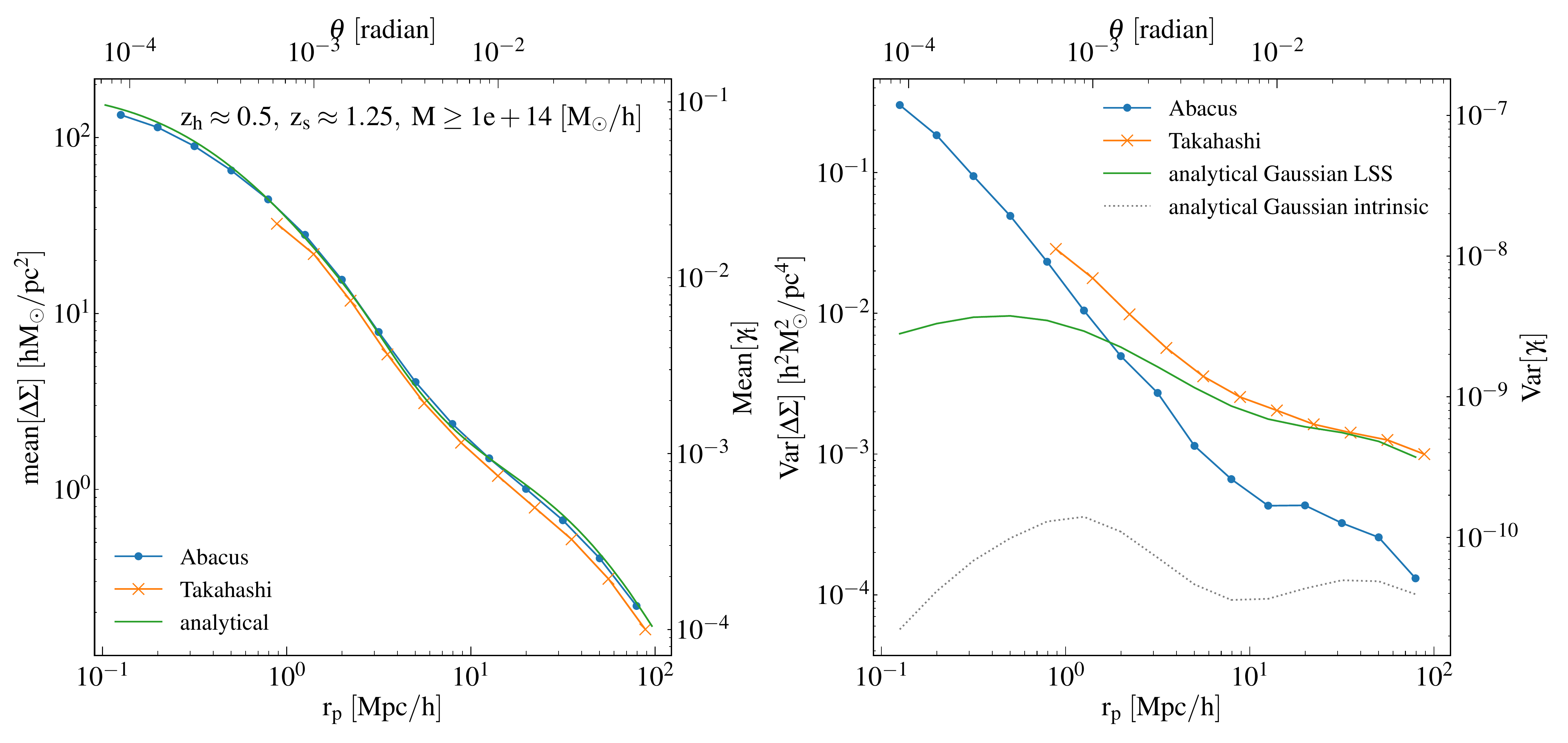}
\caption{Cluster lensing signal and variance from analytical Gaussian-field calculations (green), the Takahashi simulations (ray-tracing lensing maps, plotted in orange), and the Abacus simulations (N-body boxes, plotted in blue). 
{Left:} the mean lensing signal. The Abacus results using dark matter particles within a $\pm100\ \hiMpc$ projection depth are consistent with the ray-tracing results from Takahashi and the analytical calculations. {Right:} the variance.  The analytical calculations capture the large-scale variance (green) but not the small-scale variance (grey dotted).  On the other hand, the Abacus calculations include small-scale variance but are insufficient to account for the large-scale variance.  Therefore, we need to combine the small-scale variance from Abacus with the large-scale variance from analytical calculations.  We truncate the Takahashi curves at 1 $\hiMpc$ because of the limited resolution.}
\label{fig:abacus_takahashi}
\end{figure*}

Equation~(\ref{eq:cov_gammat}) can be interpreted as the contribution from three components: shape noise (involving $\sigma_\gamma$), LSS (involving $\clkk$), and the intrinsic variation of halo density profiles (involving $\clhk$).
\beq
\Cov^{\rm Gauss} = \Cov^{\rm shape} + \Cov^{\rm lss} + \Cov^{\rm intr}  \ .
\eeq
We will show, based on comparisons to simulations, that the LSS term is sufficiently accurate, while the intrinsic term must be supplemented by non-Gaussian calculations (e.g.~numerical simulations).

The contribution from shape noise is given by
\beqa
\Cov^{\rm shape} =  \frac{1}{4\uppi\fsky} \int \frac{\ell \dd\ell}{2\uppi} \hJ_2(\ell \theta_1)\hJ_2(\ell \theta_2)\\ \times
\left( \clhh+\frac{1}{\nh} \right)
\left(\frac{\sigma_\gamma^2}{\ns}\right)  \ .
\label{eq:cov_shape}
\eeqa
Here $\clhh$ and $1/\nh$ are associated with halo clustering and shot noise, respectively. When shot noise dominates the variance of halo number counts $( 1/\nh \gg \clhh)$, the off-diagonal elements approach zero because of the orthogonality of $\hJ_2$, and the shape noise reduces to the more intuitive form
\beqa
\Cov^{\rm shape}
& = \frac{\sigma_\gamma^2}{4\uppi \fsky \nh \ns  \uppi (\thetamax^2 - \thetamin^2)}   \\
& = \frac{\sigma_\gamma^2}{\mbox{total number of sources}}  \ .
\label{eq:cov_shape_one_over_ns}
\eeqa
In Appendix~\ref{app:bessel} we provide the derivation. 

For $M_h\gtrsim10^{14}\ \hiMsun$, halo clustering is comparable to shot noise (see the left-hand panel of Fig.~\ref{fig:cl_takahashi} and Fig.~\ref{fig:demo_shot_noise_vs_clustering_clhh_redshift_binsize}). 
For a higher mass threshold, both terms increase, and shot noise dominates for 
$M_h\gtrsim4\times10^{14}\ \hiMsun$ at $z=0.4$, and for
$M_h\gtrsim2\times10^{14}\ \hiMsun$ at $z=0.8$. 
The exact mass of transition slightly depends on the width of the redshift bin.

The contribution from the LSS is given by
\beq
\Cov^{\rm lss}  =  \frac{1}{4\uppi\fsky} \int \frac{\ell \dd\ell}{2\uppi} \hJ_2(\ell \theta_1)\hJ_2(\ell \theta_2)
\left(\clhh+\frac{1}{\nh}\right) \clkk   \ .
\label{eq:cov_lss}
\eeq
This contribution is basically the convergence power spectrum ($\clkk$, see \citealt{Kilbinger15} for a review) multiplied by the noise of halo number counts, which include both the shot noise ($1/\nh$) and the clustering of haloes ($\clhh$).\footnote{This equation is different from the equation 17 in \cite{Gruen15}, which ignores the clustering of lenses. This is the difference between the halo-to-halo covariance and the patch-to-patch covariance discussed in Section~\ref{sec:pitfalls}.}
Compared with shape noise, this LSS term is less sensitive to the radial bin size because both $\clkk$ and $\clhh$ decrease at high $\ell$ and only the integration over the first peak of $\hJ_2$ has significant contribution (see Fig.~\ref{fig:bessel_demo}).
At large scales, $\clkk$ is usually higher than $\sigma_\gamma^2/\ns$ (see the middle panel of Fig.~\ref{fig:cl_takahashi}), and thus $\Cov^{\rm lss} > \Cov^{\rm shape}$.

The contribution from the intrinsic variation of halo density profiles is given by
\beq
\Cov^{\rm intr} =  \frac{1}{4\uppi\fsky} \int \frac{\ell \dd\ell}{2\uppi} \hJ_2(\ell \theta_1)\hJ_2(\ell \theta_2)
\left(\clhk\right)^2  \ .
\label{eq:cov_intr}
\eeq
This equation assumes that matter inside haloes follows a Gaussian random field; however, the intrinsic variation of halo density profiles includes significant non-Gaussian contributions.  In this work we use N-body simulations to account for both Gaussian and non-Gaussian contributions.

Fig.~\ref{fig:abacus_takahashi} displays the mean and variance of lensing calculated from simulations and from analytical formulae.  
We use the fiducial calculations specified in Section~\ref{sec:takahashi} 
($M_{\rm 200m} \ge 10^{14} \ \hiMsun$;
$0.508 \le \zh \le 0.574$ and
$1.218 \le \zs \le 1.318$ for Takahashi; $\zh=0.5$ for Abacus), with no shape noise.  We convert from $\gamma_t$ to $\DS$ by calculating the $\Sigcrit$ corresponding to the source and lens redshifts; we show $\rp$ and $\DS$ at left and bottom axes, and the equivalent $\theta$ and $\gamma_t$ on the top and right axes.

The left-hand panel shows the mean value of $\DS$ calculated analytically from equation~(\ref{eq:DS_realspace}).  The right-hand panel shows the contributions from LSS ($\Cov^{\rm lss}$, green) and from the Gaussian intrinsic term ($\Cov^{\rm intr}$, grey dotted).  The LSS contribution fully accounts for the variance above $10^{-2}$ radian ($\approx$ 10 $\hiMpc$).  The Gaussian intrinsic term significantly under-predicts the small-scale variance and will be replaced by N-body simulations (Section~\ref{sec:grafting}).

We do not include shape noise in this figure. When halo clustering is negligible, the shape noise is inversely proportional to the area of the radial bin and the surface density of sources (equation~\ref{eq:cov_shape_one_over_ns}).  At small scales, the shape noise is usually higher than the intrinsic variation of the halo density profile.  At large scales, the shape noise becomes subdominant to the contribution from LSS, and the transition scale depends on the source density.  We will discuss the relative importance of shape noise in detail in Section~\ref{sec:shape_noise}.  For the high source densities of LSST, {\rm Euclid}, and {\rm WFIRST}, the small-scale intrinsic variation of halo density profiles will become non-negligible, and it is imperative to accurately characterise the covariance at small scales.

\subsection{Angular power spectra}\label{sec:cl}

\begin{figure*}
\centering
\includegraphics[width=1.0\textwidth]{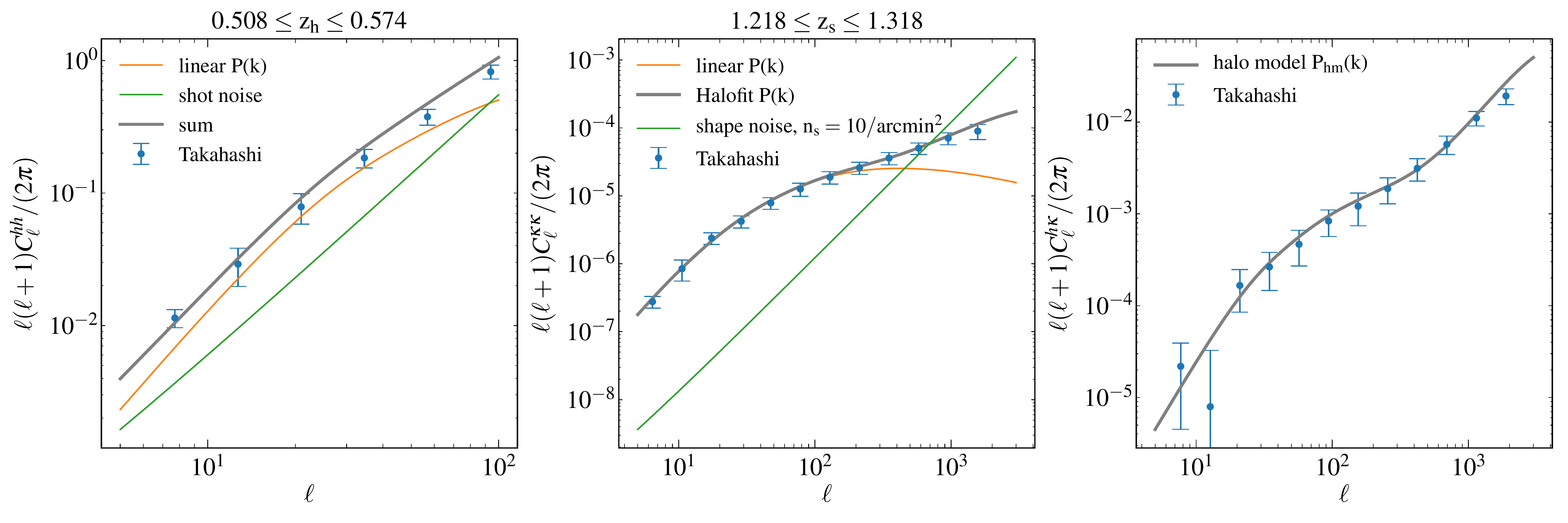}
\caption{Angular power spectra of lenses and sources from the Takahashi simulations (blue) and from analytical calculations (grey).  {Left}: auto power spectrum of the number overdensity of haloes.  The simulation agrees with the sum of halo clustering (equation~\ref{eq:clhh}, orange) and shot noise ($1/\nh$, green).  {Middle}: auto power spectrum of convergence. The simulation agrees with equation~(\ref{eq:clkk}) when using the {\sc Halofit} non-linear matter power spectrum.  Shape noise dominates at small scales (green).  {Right}: cross power spectrum between halo and convergence.  The simulation agrees with equation~(\ref{eq:clhk}) based on the halo-matter cross power spectrum from the halo model.}
\label{fig:cl_takahashi}
\end{figure*}

We use the Limber approximation to calculate the angular power spectra in equation~(\ref{eq:cov_gammat}).  In Appendix~\ref{app:limber} we provide the derivations (also see \citealt{LoverdeAfshordi08}). 

The auto angular power spectrum of the halo number density is given by 
\beq
\clhh = \int_{\chimin}^{\chimax} \dd\chih \left(\frac{F_{\rm h}(\chih)}{\chih} \right)^2
P_{\rm hh}\left(k=\frac{\ell{+1/2}}{\chih}\right) \ ;
\label{eq:clhh}
\eeq
the $\chih$ integration is over the comoving distance range ($\chimin$, $\chimax$) of the cluster redshift bin, and
\beq
F_{\rm h}(\chih) = \frac{\chih^2}{V}; \quad V=\int_{\chimin}^{\chimax} \chih^2 \dd\chih  \ .
\label{eq:F_h}
\eeq
For $P_{\rm hh}$,
we use the linear matter power spectrum multiplied by the halo bias from \cite{Tinker10}.  
The left-hand panel of Fig.~\ref{fig:cl_takahashi} shows the $\clhh$ calculated from equation~(\ref{eq:clhh}) and from the Takahashi simulations, which agree with the sum of $\clhh$ and the shot noise $1/\nh$.  
In Appendix~\ref{app:limber} we show that wider redshift bins correspond to lower $\clhh$ and $1/\nh$, while the relative importance of the two remains unchanged.

For the convergence power spectrum, we assume that the source galaxies follow a redshift distribution $\psrc(\chis)$ normalised such that $\int_0^{\infty} \dd\chis \psrc(\chis) = 1$.  The auto angular power spectrum of convergence is given by 
\beq
\clkk = \int_0^{\infty} \dd\chiLSS  \left( \frac{F_\kappa(\chiLSS)}{\chiLSS}\right)^2\Pmm\left(k=\frac{\ell{+1/2}}{\chiLSS}\right) \ ,
\label{eq:clkk}
\eeq
where we integrate all the LSS along the line of sight, and 
\beq
F_\kappa(\chiLSS) = \bar{\rho}
\int_{\chiLSS}^\infty  \dd\chis 
\psrc(\chis)\frac{1}{\Sigcrit(\zs,\zLSS)}  \ .
\label{eq:F_kappa}
\eeq
Equation~(\ref{eq:clkk}) corresponds to the contribution of lensing from all the LSS in front of the sources, integrated from zero distance to the farthest source galaxy.
Equation~(\ref{eq:F_kappa}) shows that the intervening LSS
is weighted by the lensing kernel.  Additional weights on sources can be absorbed in $\psrc$.  This equation is equivalent to the equation 29 in \cite{Kilbinger15}.

The middle panel of Fig.~\ref{fig:cl_takahashi} shows the $\clkk$ calculated from equation~(\ref{eq:clkk}) and from the Takahashi simulations.  In the analytical calculation, the linear matter power spectrum underestimates the small-scale (high-$\ell$) power, while the non-linear matter power spectrum from {\sc Halofit} \citep{Smith03} agrees with simulations out to $\ell\sim1000$.
We also show the shape noise term $\sigma_\gamma^2/\ns$ for $\sigma_\gamma=0.3$ and $\ns$=10 ${\rm arcmin}^{-2}$; the shape noise dominates at small scale.

The cross angular power spectrum of halo and lensing is given by
\beq
\clhk = \int_{\chimin}^{\chimax} \dd\chih  \bigg( \frac{F_\kappa(\chih)}{\chih} \bigg) 
\bigg( \frac{F_{\rm h}(\chih)}{\chih} \bigg)
P_{\rm hm}\left(k=\frac{\ell{+1/2}}{\chih}\right)  \ .
\label{eq:clhk}
\eeq
Similar to $\clhh$, this integration is over the line-of-sight distance range of the lens sample.  The right-hand panel of Fig.~\ref{fig:cl_takahashi} shows the $\clhk$ calculated from equation~(\ref{eq:clhk}) and from the Takahashi simulations.  We use the 3D halo--matter cross power spectrum $P_{\rm hm}$ from the halo model; i.e., the sum of the two-halo term (linear power spectrum multiplied by halo bias) and the one-halo term computed for an NFW density profile (see Appendix~\ref{app:halo_model}).

\vspace{-7mm}
\section{Analytical Gaussian-field covariance for excess surface density ($\DS$)}\label{sec:cov_DS}

In this section, we focus on the covariance of the excess surface density ($\DS$).  Since $\DS$ and $\gamma_t$ are proportional to each other, their fractional errors are the same. When we consider a single source redshift (that is, if $\psrc$ is close to a delta function), the covariance matrices of $\gamma_t$ and $\DS$ simply differ by a constant $\Sigcrit^2(\zs, \zh)$.  However, when we consider a broad range of source redshifts, we need to integrate $\psrc(\zs)\Sigcrit(\zs, \zh)/\Sigcrit(\zs, \zLSS)$ and change the order of integration; see equation~(\ref{eq:F_Sigma}) below.  Therefore, we find it necessary to detail the analytical expressions for the covariance of $\DS$.  

The contribution of LSS to the covariance of $\DS$ is less intuitive than that of $\gamma_t$, because it is the projected mass density weighted by the lensing kernel and is equivalent to taking the $\gamma_t$ coming from large scale structure between redshift 0 and $\zs$ and treating it as if it is at $\zh$. This component is not the projected mass density of LSS because the thin-lens approximation does not hold for LSS.

The covariance matrix analogous to equation~(\ref{eq:cov_gammat}) is given by\footnote{For a single source redshift, equation~(\ref{eq:cov_DS}) is equivalent to the Gaussian-field part of equation A40 in \cite{Singh17}.  We use $C_\ell$'s to make the equations analogous to equation~(\ref{eq:cov_gammat}), while \cite{Singh17} uses 2D power spectra.}
\beqa
\Cov^{\rm Gauss}[\DS(\rp), \DS(\rp')] = \frac{1}{4\uppi f_{\rm sky}}\int \frac{k \dd k}{2\uppi} \hJ_2(k \rp) \hJ_2(k \rp') \times\\ 
\left[
 \left(\clhh + \frac{1}{\nh}\right)
\left(\clSS + \big\langle{\Sigcrit}\big\rangle^2\frac{\sigma_\gamma^2}{\ns}\right)
+ \left(\clhS\right)^2 
\right] \ ,
\label{eq:cov_DS}
\eeqa
where
\beq
\big\langle{\Sigcrit}\big\rangle = \int_0^\infty \dd\chis \psrc(\chis)\Sigcrit(\zs, \zh) \ .
\eeq
To derive this equation, we use the covariance of $\gamma_t$ (equation~\ref{eq:cov_gammat}) for a single source redshift and a single lens redshift, and then integrate over $\int \dd\chis \psrc(\chis)\Sigcrit(\zs, \zh)$ twice.  To convert from ($\ell$, $\theta$) to ($k$, $\rp$), we assume  $\theta = \rp/ \chih $  and $\ell = k \chih $; therefore, this expression only applies to a thin lens redshift bin.

Here we introduce two extra angular power spectra: $\clSS$ corresponds to the auto spectrum for projected matter (analogous to $\clkk$), and $\clhS$ corresponds to the cross spectrum between halo and projected matter (analogous to $\clhk$). Using the Limber approximation (see Appendix~\ref{app:limber}),
the auto power spectrum for projected matter is given by
\beq
\clSS(\zh) = \int_0^\infty \dd\chiLSS  \left(\frac{F_{\Sigma}(\chiLSS, \chih)}{\chiLSS} \right)^2 \Pmm\left(k=\frac{\ell{+1/2}}{\chiLSS}\right) \ .
\label{eq:clSS}
\eeq
Here we integrate the LSS along the line-of-sight from zero to infinity and weight the LSS by the window function 
\beq
F_\Sigma(\chiLSS,\chih) =  \bar{\rho} \int_{\chiLSS}^{\infty} \dd\chis \psrc(\chis)   \frac{\Sigcrit(\zs, \zh)}{\Sigcrit(\zs, \zLSS)}  \ .
\label{eq:F_Sigma}
\eeq
The $\Sigcrit(\zs, \zLSS)$ in the denominator comes from the lensing kernel (the same as in equation~\ref{eq:F_kappa}), while the $\Sigcrit(\zs, \zh)$ in the numerator comes from the fact that we interpret all the line-of-sight structure as the noise to halo profiles at the haloes' redshift $\zh$.

The halo-matter cross power spectrum is given by
\small
\beq
\clhS(\zh) = 
\int_{\chimin}^{\chimax} \dd\chi
\left(\frac{F_{h}(\chi)}{\chi} \right)
\left(\frac{F_{\Sigma}(\chi,\chih)}{\chi} \right)  \Phm\left(k=\frac{\ell{+1/2}}{\chi}\right) \ .
\label{eq:clhS}
\eeq
\normalsize
Here we integrate over the redshift range of the halo sample (where the two fields $h$ and $\Sigma$ overlap), and $F_{\rm h}$ is given by equation~(\ref{eq:F_h}).  Here $\clSS$ has the same dimension as $\Sigma^2$  and $\clhS$ has the same dimension as $\Sigma$.

\section{Small-scale covariance from N-body simulations}\label{sec:grafting}

When shape noise is subdominant, Fig.~\ref{fig:abacus_takahashi} shows that equation~(\ref{eq:cov_gammat}) underestimates the covariance at small scales, a consequence of treating matter in haloes as a Gaussian random field.  The actual covariance in this regime will include the effects of variation in halo concentration, sub-structures, and orientations, none of which are captured by the Gaussian field approximation.  Similar to \cite{Gruen15}, we calculate  cluster lensing using N-body simulations.


\subsection{Abacus N-body simulations}\label{sec:abacus}

We use the publicly available Abacus Cosmos simulations\footnote{ \href{https://lgarrison.github.io/AbacusCosmos/}{https://lgarrison.github.io/AbacusCosmos/}.} \citep{Garrison18} based on the {\sc Abacus} N-body code \citep{Metchnik09,Garrison18}.
We use the 20 realisations of {\tt AbacusCosmos\_720box\_planck}.  This suite of boxes are based on a \cite{Planck15Cosmo} cosmology ($\Omega_m$=0.314, $\sigma_8=0.83$, $h$=0.67) with different phases in the initial condition.  
Each realisation has 1440$^3$ particles in a box of side length 720 $\hiMpc$, 
a mass resolution of $1\times10^{10}\ \hiMsun$, and
a spline softening length of 41 $\hikpc$.
Dark matter haloes are identified using {\sc Rockstar} \citep{Behroozi13rs}.

For computing the covariance matrices, we divide each box in the $x$-$y$ plane into 9 equal prism-shaped subvolumes, each of which has a dimension of $80\times80\times720\ (\hiMpc)^3$.  Each subvolume includes approximately 580 haloes with $M_{\rm 200m}\ge 10^{14}\ \hiMsun$.  We calculate the covariance matrix from these 20 $\times$ 9 = 180 subvolumes.  Since the covariance is inversely proportional to the simulation volume, we multiply the covariance by the volume in the unit of $(\hiGpc)^3$ so that the resulting covariance matrix corresponds to a 1 $(\hiGpc)^3$ volume.

We use the 10\% down-sampled particles to measure the azimuthally averaged $\DS$ profiles around haloes.  We use an integration depth of $\pm 100\ \hiMpc$ along the $z$-direction of the box; this integration depth is sufficient for a convergent $\DS$ profile but insufficient to include the contribution of uncorrelated LSS to the covariance matrix.  We will use analytical calculations to capture the contribution from the LSS outside this $\pm100\ \hiMpc$ integration depth.

To compute $\DS$, we cross-correlate the haloes in a subvolume with the particles in the full-volume, applying the periodic boundary condition. For haloes near the boundary of each subvolume, we use particles outside the boundary to measure their $\DS$.\footnote{Since we are calculating cross correlation between haloes and dark matter particles, this treatment is not affected by boundary-pair problems pointed out by \cite{Friedrich16}.}
For counting halo-particle pairs, we use the public code {\sc Corrfunc}\footnote{
\href{https://github.com/manodeep/Corrfunc}{https://github.com/manodeep/Corrfunc}} \citep{Corrfunc}.
Similar to the $\gamma_t$ calculation, we calculate $\DS$ around random points (30 times the number of haloes) in each subvolume and subtract it from the $\DS$ around haloes.

Fig.~\ref{fig:abacus_takahashi} compares the results from the Abacus simulations with the Takahashi simulations and with analytical calculations.  The left-hand panel shows the mean $\DS$.  For Abacus, we use the $z=0.5$ output.  For Takahashi, we use the lens redshift $0.508 \le \zh \le 0.574$ and source redshift $1.218 \le \zs \le 1.318$, converting from $\gamma_t$ to $\DS$ using the corresponding $\Sigcrit(\zs, \zh)$.   The slight difference of the two simulations is due to the slightly different redshift and cosmology ($\Omega_m$=0.314 for Abacus and 0.279 for Takahashi).  Because of the relatively low resolution of Takahashi, the density profile is underestimated below 1 $\hiMpc$.  We calculate the mean profile analytically using equation~(\ref{eq:DS_realspace}) using $\xihm$ from the halo model (assuming the same cosmology as the Abacus boxes we use).

The right-hand panel of Fig.~\ref{fig:abacus_takahashi} shows the variance of $\DS$.  The Takahashi result is significantly higher than the Abacus result at large scales because the former includes the lensing effects from all the LSS, while the latter only takes into account $\pm$100 $\hiMpc$. At intermediate scales, the Takahashi result approaches the Abacus result because the intrinsic variation of halo density profiles starts to dominate.

\subsection{Combining analytical and numerical treatments}

\begin{figure}
\centering
\includegraphics[width=0.5\textwidth]{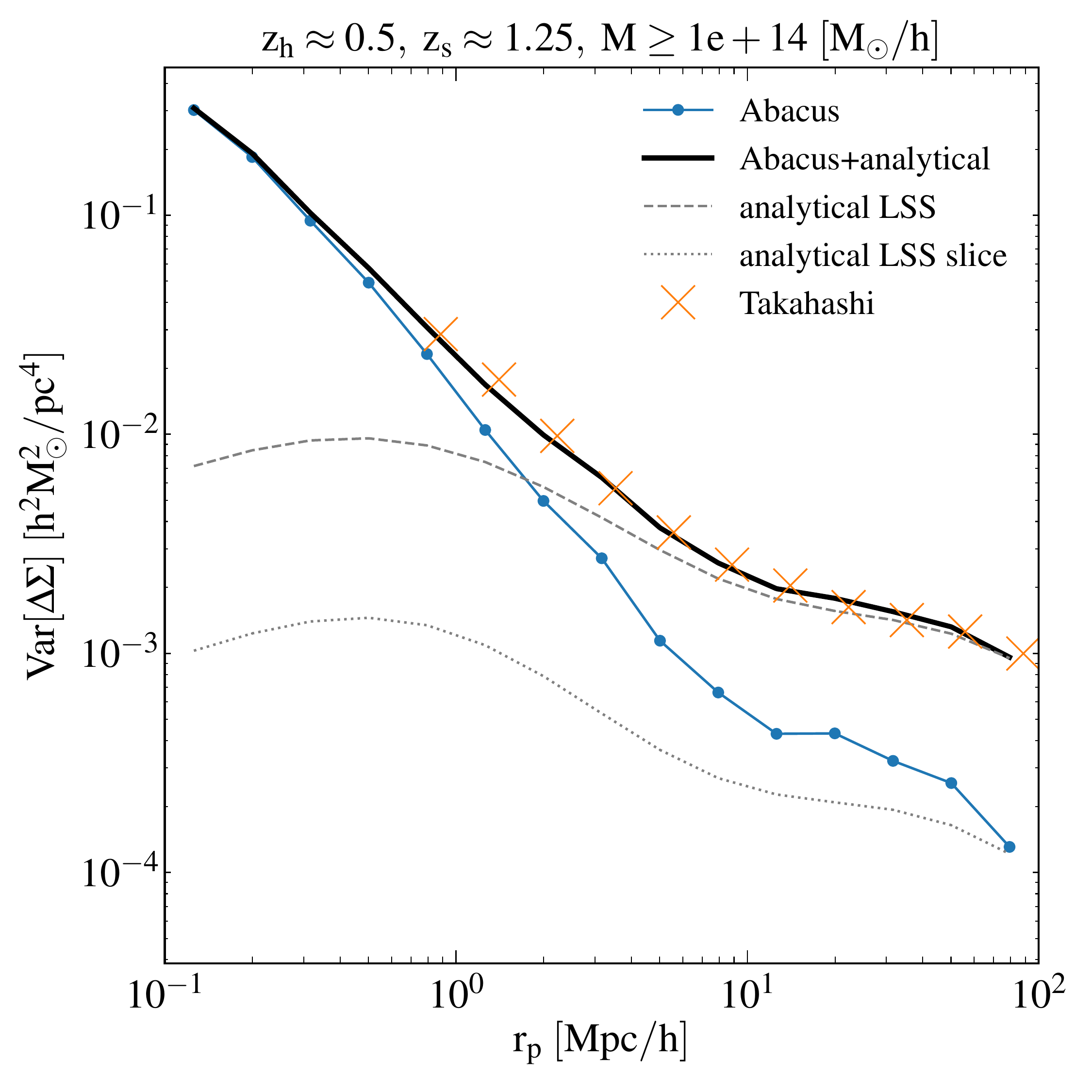}
\caption{Variance combining the Abacus simulations and analytical calculations for negligible shape noise. The small-scale variance (dominated by intrinsic variation of halo density profiles) comes from the Abacus simulations (blue), while the large-scale variance (dominated by LSS) comes from analytical calculations (grey dash).  The simulation results include a slice of LSS near haloes, which can be calculated analytically (grey dotted) and is subtracted from the full LSS calculation. The combined results (black) agree well with the Takahashi ray-tracing simulations (orange crosses).}
\label{fig:abacus_analytic_combined}
\end{figure}

Our approach to a full calculation of covariance matrices is to combine the analytical expressions for shape noise and LSS contributions (equations~\ref{eq:cov_shape} and \ref{eq:cov_lss}) with Abacus calculations, which model the intrinsic halo profile contribution and replace the inaccurate Gaussian-field model at small scales (equation~\ref{eq:cov_intr}).  However, our Abacus calculations also include the LSS contribution from a $\pm$100 $\hiMpc$ slice.  We calculate the contribution from this slice by integrating $\clkk$ or $\clSS$ from $\chi-\Delta\chi$ to $\chi+\Delta\chi$, with $\chi$ corresponding to the halo redshift and $\Delta\chi=100\ \hiMpc$.  We then subtract this slice from the full LSS contribution.

Fig.~\ref{fig:abacus_analytic_combined} shows how we combine small-scale simulation results with large-scale analytical results.  At small scales, the Abacus calculation dominates; at large scale, the Abacus calculation is similar to but slightly higher than the Gaussian-field LSS in a slice of $\pm100\ \hiMpc$ (grey dotted).  This difference could be related to the correlated structure within the slice.  We subtract this slice of LSS from the full LSS and add the Abacus variance.  In practice, subtracting this slice has negligible effect.  The combined result is the heavy black curve. We also show the results from the Takahashi shear maps;  we use the $\gamma_t$ measurements from Takahashi in a narrow lens range ($ 0.508\le \zh \le 0.574$) and a narrow source range ($1.218 \le \zs \le 1.318$), calculate the corresponding $\Sigcrit$, and scale to a 1 $(\hiGpc)^3$ volume.  The Takahashi results agree well with our Abacus+analytical approach at intermediate and large scales.  At small scales, the Takahashi simulation cannot resolve the inner profile of clusters, which has been shown in Fig.~\ref{fig:abacus_takahashi}.

We have also attempted to calculate the non-Gaussian small-scale covariance analytically.  We find that the analytical non-Gaussian results are higher than simulations (see Appendix~\ref{app:non_gaussian}).  This discrepancy does not affect any of the calculations in the main text.

\section{Correlation between radial bins}\label{sec:off-diagonal}
\begin{figure*}
\centering
\includegraphics[width=\textwidth]{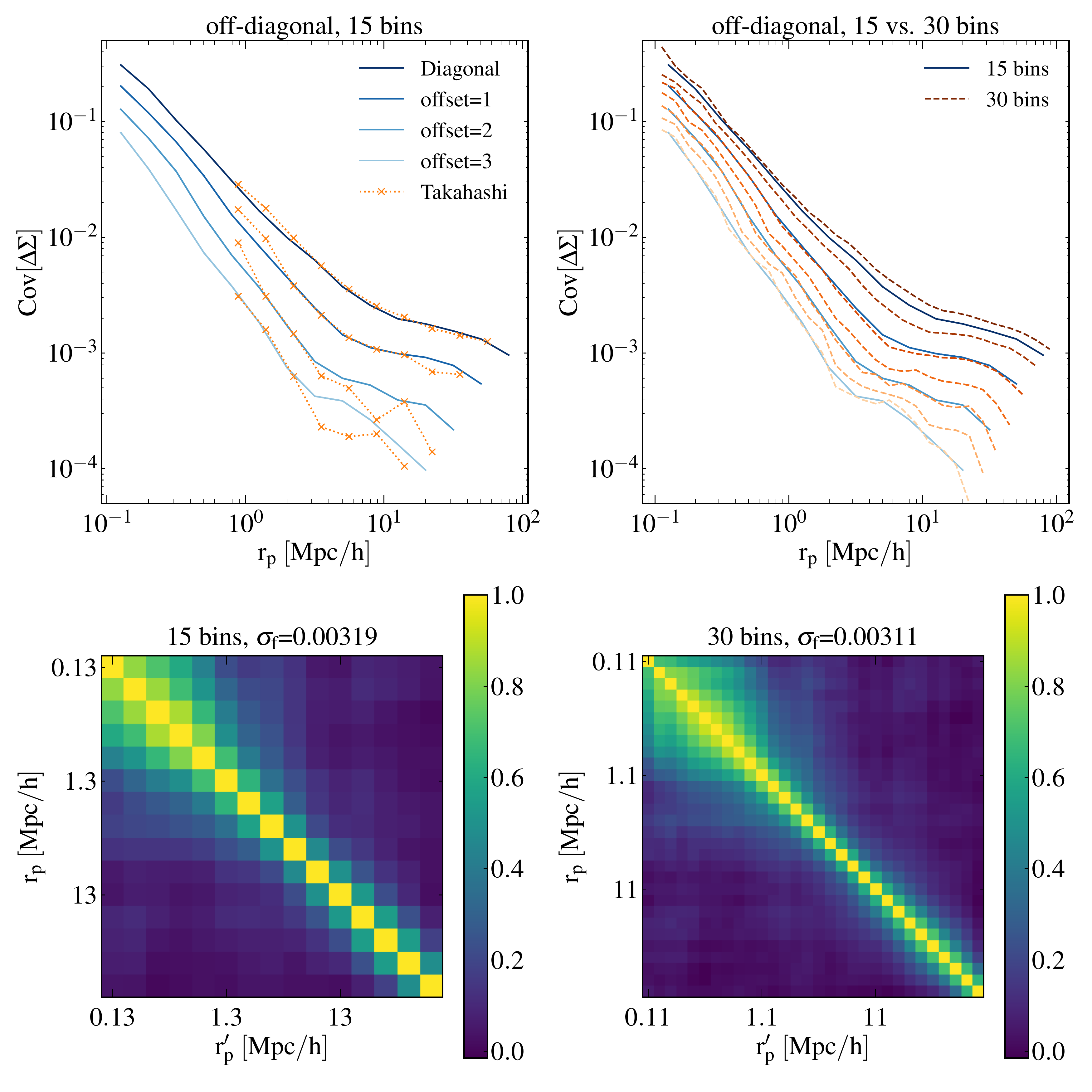}
\caption{Covariance matrices for 15 or 30 logarithmically-spaced radial bins between 0.1 and 100 $\hiMpc$.
{Top left:} matrix elements for 15 radial bins. From the top curve: 15 elements on the diagonal, 14 elements next to the diagonal, and so on. The blue curves are from Abacus+analytical, while the orange curves are from Takahashi simulations.
{Top right:}
comparison between 15 bins and 30 bins. For 30 bins (orange dash), every other curve agrees with a curve of 15 bins (blue solid), indicating that at a given $(\rp, \rp')$, the covariance is approximately independent of the bin size. 
{Bottom:}
correlation matrices for 15 radial bins (bottom left) and 30 radial bins (bottom right).  The structures of the two matrices are almost identical if we ignore the different bin sizes.
}
\label{fig:off_diagonal_radial_binsize}
\end{figure*}
\begin{figure*}
\includegraphics[width=0.49\textwidth]{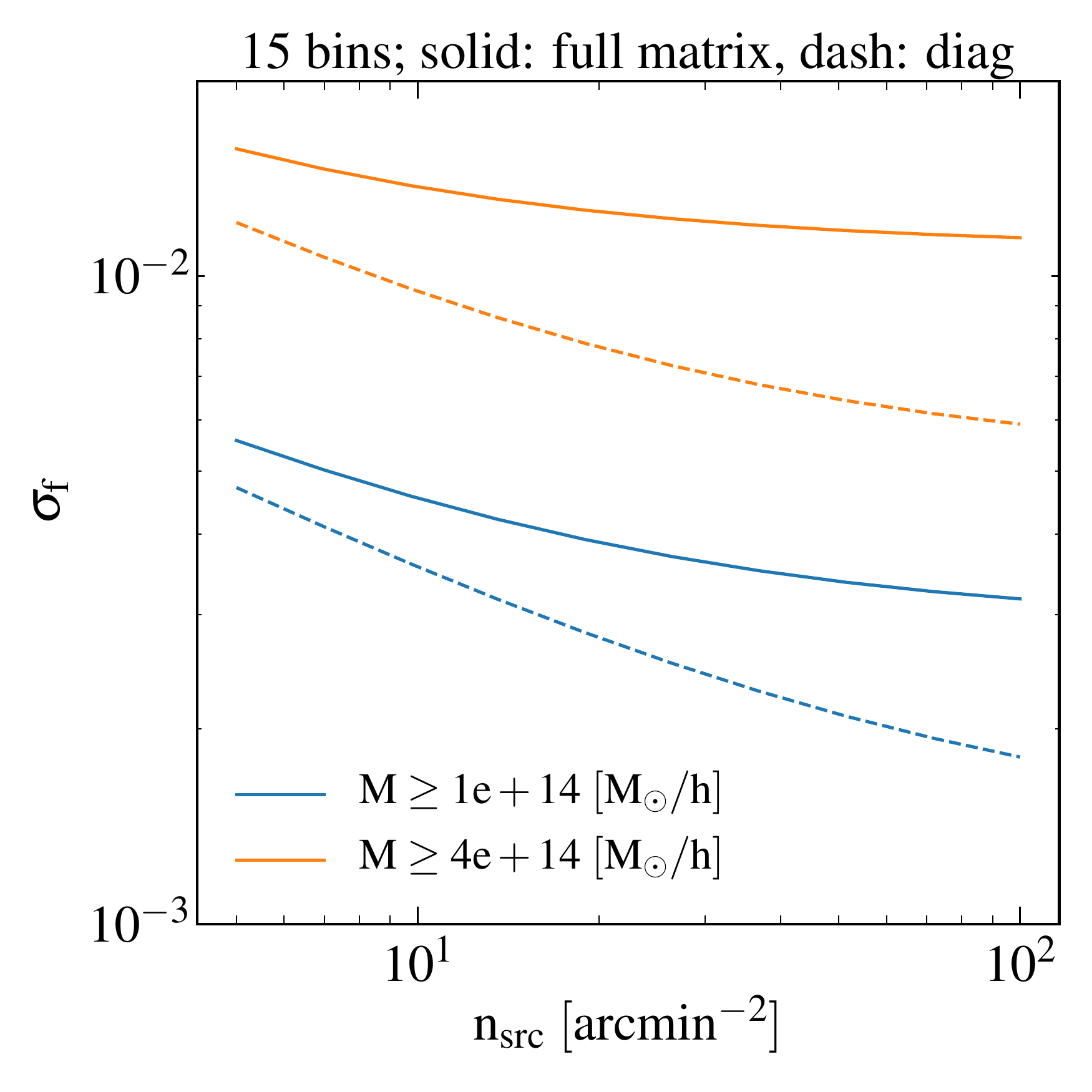}
\includegraphics[width=0.49\textwidth]{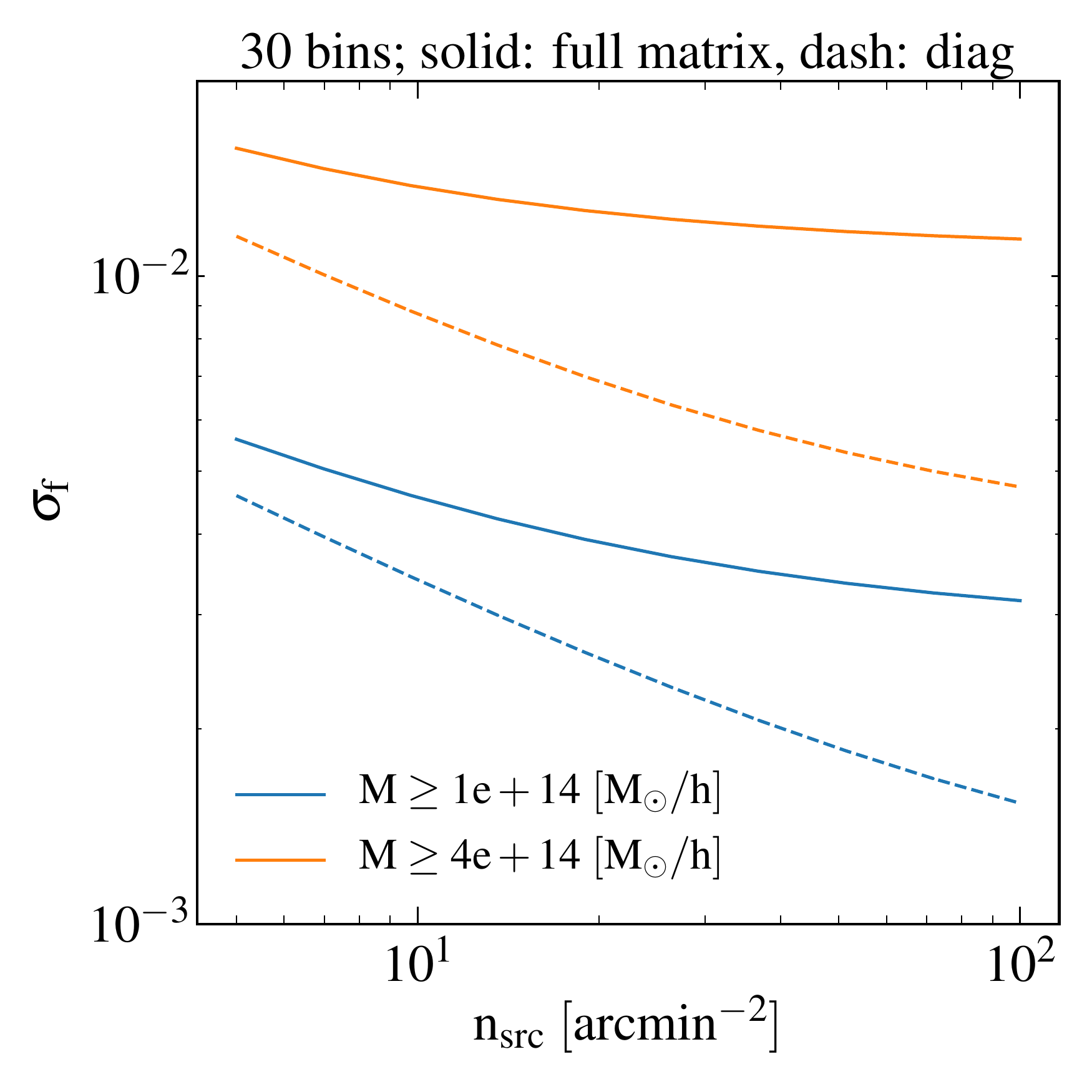}
\caption{Impact of bin size, shape noise, mass threshold, and off-diagonal elements on the constraints of a multiplicative parameter $\sigmaf$ (equation~\ref{eq:sigmaf}), for a survey of 5000 deg$^2$, $0.4\le z<0.6$.  The two panels correspond to 15 and 30 radial bins.  The $x$-axis corresponds to various source densities.  The solid curves correspond to using the full covariance matrices, while the dash curves correspond to using only the diagonal elements.  The former is independent of bin size, confirming that the information content is independent of bin size.  The latter underestimates $\sigmaf$, and the underestimation is worse for narrower bins, higher halo mass, and higher source densities.}
\label{fig:constraints_vs_nsrc_5000sqdeg}
\end{figure*}
\begin{figure*}
\includegraphics[width=0.49\textwidth]{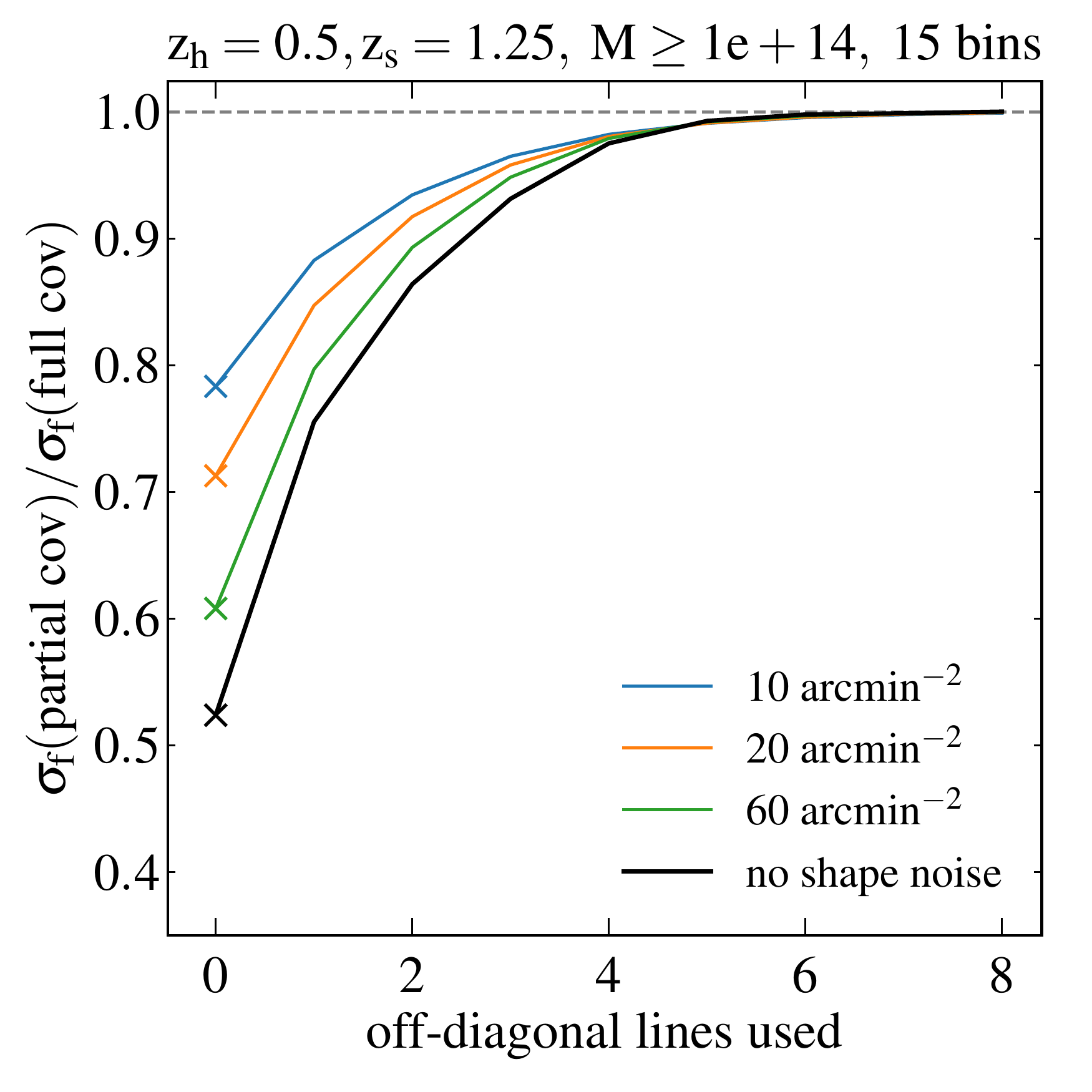}
\includegraphics[width=0.49\textwidth]{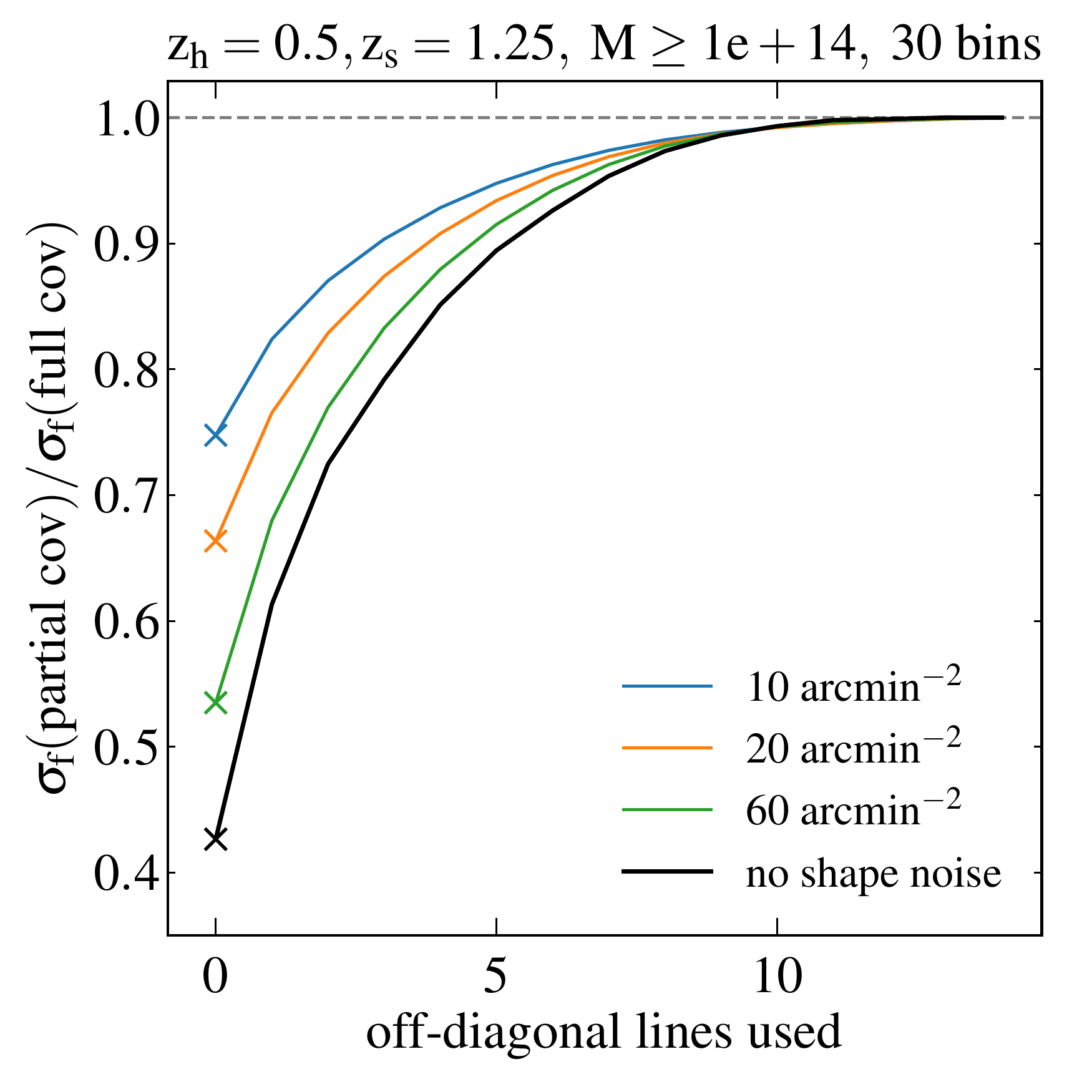}
\caption{Impact of off-diagonal elements on  the constraints of a multiplicative parameter $\sigma_f$ (equation~\ref{eq:sigmaf}), calculated with the full covariance matrix vs.\ a partial covariance matrix (using elements close to the diagonal).  The $x$-axis corresponds to the number of off-diagonal lines used: $x=0$ indicates using diagonal elements only, $x=1$ indicates including the elements next the diagonal, and so on.  The $y$-axis corresponds to $\sigmaf$ for each case divided by the $\sigmaf$ using the full covariance matrix.  Different curves correspond to different assumptions of shape noise.  In the absence of shape noise, using only the diagonal elements would lead to a factor of two underestimation of $\sigmaf$.
{Left:} 15 radial bins.  Using 3 off-diagonal lines ($x=3$) is very close to using the full covariance matrix.  With larger shape noise, fewer off-diagonal lines are needed.  {Right:} 30 radial bins. In this case, we need to double the number of off-diagonal lines used compared with the 15-bin case.}
\label{fig:constraints_off_diag}
\end{figure*}

In this section, we discuss the properties of full covariance matrices, focusing on the importance of off-diagonal elements.  Fig.~\ref{fig:off_diagonal_radial_binsize} shows the full covariance matrices corresponds to  Fig.~\ref{fig:abacus_analytic_combined}:
$M_{\rm 200m}\ge 10^{14}\ \hiMsun$,
$\zh=0.5$,
$\zs=1.25$, with no shape noise.
We combine Abacus simulations and analytical calculations, and we use 15 or 30 logarithmically-spaced $\rp$ bins between 0.1 and 100 $\hiMpc$.

The top panels show the diagonal elements and the off-diagonal elements parallel to the diagonal. The top curves correspond to the diagonal elements, the second curves correspond to the elements next to the diagonal (offset from the diagonal by 1 element), and so on.  In the top left panel, we show both the Abacus+analytical (blue) and Takahashi (orange crosses) results for 15 radial bins.  The values of the off-diagonal elements drop by approximately an order of magnitude at the fourth curve (offset=3), which corresponds to a 0.6 dex difference between $\rp$ and $\rp'$. The off-diagonal elements drop more rapidly at large scales than at small scales because small-scale structures are more correlated.

The top right panel compares the elements for 15 radial bins (blue solid) and 30 radial bins (orange dash).  For the diagonal elements, the dependence on bin size is rather weak.  This is explained by the property of the bin-averaged Bessel function $\hJ_2$ (see Appendix~\ref{app:bessel} and Fig.~\ref{fig:bessel_demo}).  The peak of $\hJ_2$ is insensitive to the radial bin size; given the steep negative slope of $\clkk$ and $\clhh$ at large $\ell$, only the first peak of $\hJ_2$ contributes to the integration in equation~(\ref{eq:cov_gammat}), and thus the dependence on the bin size is weak.  This is contrary to the case of shape noise, which is inversely proportional to the bin area and only contributes to the diagonal elements when halo shot noise dominates.  The insensitivity to the radial bin size at small scales in the N-body results also indicates that the particle shot noise is negligible; if we have a low number of dark matter particles in N-body simulations, the small-scale variance would be inversely proportional to the number of particles in a radial bin.

For the 30-bin case (orange dash), the first, third, and fifth curves agree with the first, second, and third curves of the 15-bin case (blue solid).  That is, every other curve of the 30-bin case agrees with the 15-bin case, and each of the even-numbered orange dash curves is approximately the geometric mean of the two neighbouring curves.  In other words, for a given ($\rp$, $\rp'$), the covariance is almost independent of the bin size.  By halving the bin size, we increase the number of correlated bins, while the covariance between the two scales remains the same.

The two bottom panels show the correlation matrices (with the diagonal elements normalised to 1), 
\beq
\rm Cor_{ij}=\frac{Cov_{ij}}{\sqrt{Cov_{ii}Cov_{jj}}} \ ,
\eeq
for 15 and 30 radial bins, respectively.  
The two matrices have similar structures apart from the different bin sizes.

To compare the information content of these matrices, we assume a parameter $f$ multiplying the $\DS$ data vector. The Fisher information for $f$ is given by
\beq
F = \rm \DS^T \left(Cov^{-1}\right) \DS \ ,
\eeq
where the superscript $T$ denotes the transpose.  This expression is equivalent to the square of the signal-to-noise ratio.  The constraint on $f$ is given by
\beq
\sigmaf = F^{-1/2}  
\label{eq:sigmaf}
\eeq
and corresponds to the constraint on the lensing amplitude.
The $\sigmaf$ values from the two matrices are almost identical (0.0032 and 0.0031), confirming that the information content is independent of the radial bin size.

Fig.~\ref{fig:constraints_vs_nsrc_5000sqdeg} further demonstrates the impact of bin size, shape noise, and off-diagonal elements on $\sigmaf$.  We assume a survey area of 5000 deg$^2$ and a redshift range $0.4\le z < 0.6$, which corresponds to a comoving volume of 1.2 $(\hiGpc)^3$.  We perform the calculation for two mass thresholds: $10^{14}$ and $4\times10^{14}\ \hiMsun$.  The left- and right-hand panels correspond to 15 and 30 radial bins, respectively.  The $x$-axis corresponds to different levels of shape noise indicated by the source density. 
The solid curves correspond to $\sigmaf$ calculated using full covariance matrices, while the dash curves use only the diagonal elements and underestimate $\sigmaf$.
For the full-covariance cases (solid curves), the two panels show identical results, indicating that the total signal-to-noise ratio does not depend on the radial bin size.  Comparing the two mass ranges, we can see that the low-mass threshold is more sensitive to the shape noise than the high-mass threshold because of its larger statistical power.

For the diagonal-only cases (dash curves), the two panels have almost the same results when shape noise dominates (low source density) because most of the off-diagonal elements are negligible.  When the shape noise is low (high source density), the two panels show slightly different results, which originate from the slightly different diagonal elements for different bin sizes (see the top right panel of Fig.~\ref{fig:off_diagonal_radial_binsize}, two top curves).  In addition, the underestimation of $\sigmaf$ is worse for narrower radial bins, lower shape noises (high source densities), and higher halo mass.

Since the diagonal elements drop rapidly, the elements far from the diagonal should be negligible.  Fig.~\ref{fig:constraints_off_diag} demonstrates how many off-diagonal elements are needed to avoid underestimation of $\sigmaf$.  We compare the $\sigmaf$ using the full covariance matrix vs.\ part of the covariance matrix (ignoring elements far from the diagonal).  The two panels correspond to 15 and 30 radial bins.  The $x$-axis corresponds to the number of the off-diagonal lines used; $x=0$ corresponds to using only the diagonal elements (marked by crosses), $x=1$ corresponds to adding the first off-diagonal line, and so on. The $y$-axis corresponds to the ratio of the $\sigmaf$ calculated with the partial covariance matrix specified by the $x$-axis and the $\sigmaf$ obtained from the full matrix.  Various curves correspond to source densities 10 arcmin$^{-2}$ (similar to DES), 20, 60 (similar to {\rm WFIRST}), and $\infty$ (no shape noise).

The left-hand panel shows that for 15 radial bins, for the case with no shape noise (black curves), using only the diagonal elements ($x=0$) underestimates the error bar by a factor of 2, while including 3 off-diagonal lines ($x=3$) is almost equivalent to using the full covariance matrix.  Comparing the two panels we can see that when we halve the bin size, we need to double the number of off-diagonal lines we use.  When the source density is low (shape noise is high), the effect of off-diagonal elements is weaker but still non-negligible.  Even with 10 arcmin$^{-2}$, we cannot completely ignore the off-diagonal elements because the shape noise is subdominant at large scales.  Other redshifts and mass thresholds give very similar results.

\section{Discussion}\label{sec:discussion}

In this section, we discuss the importance of shape noise (\ref{sec:shape_noise}), the mass and redshift dependence of covariance matrices (\ref{sec:M_z_dependence}), the cross-covariance of two different mass bins (\ref{sec:cov_cross_mass}), and potential systematic uncertainties that can affect our calculations (\ref{sec:systematics}).

\subsection{Importance of shape noise}\label{sec:shape_noise}

\begin{figure*}
\includegraphics[width=0.48\textwidth]{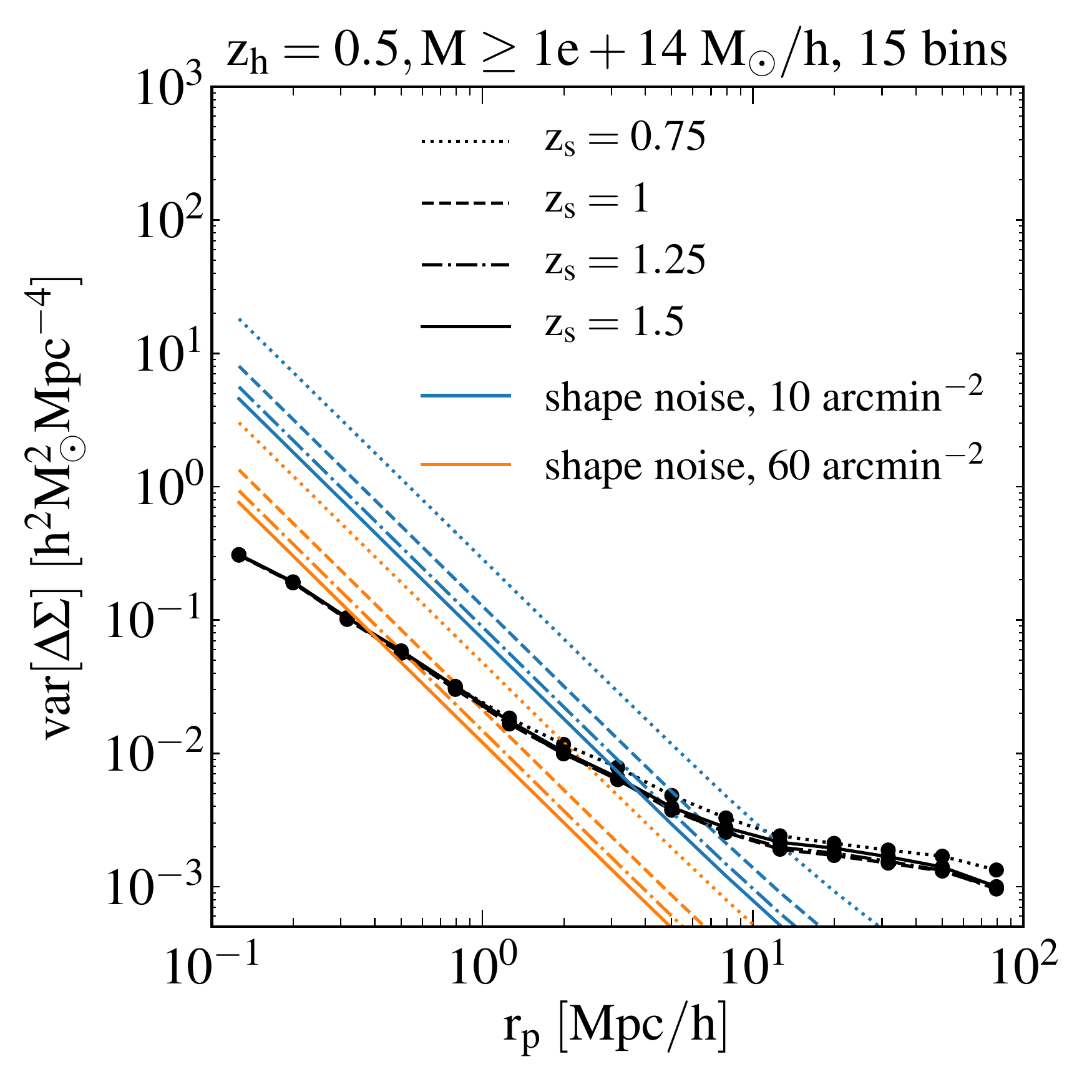}
\includegraphics[width=0.48\textwidth]{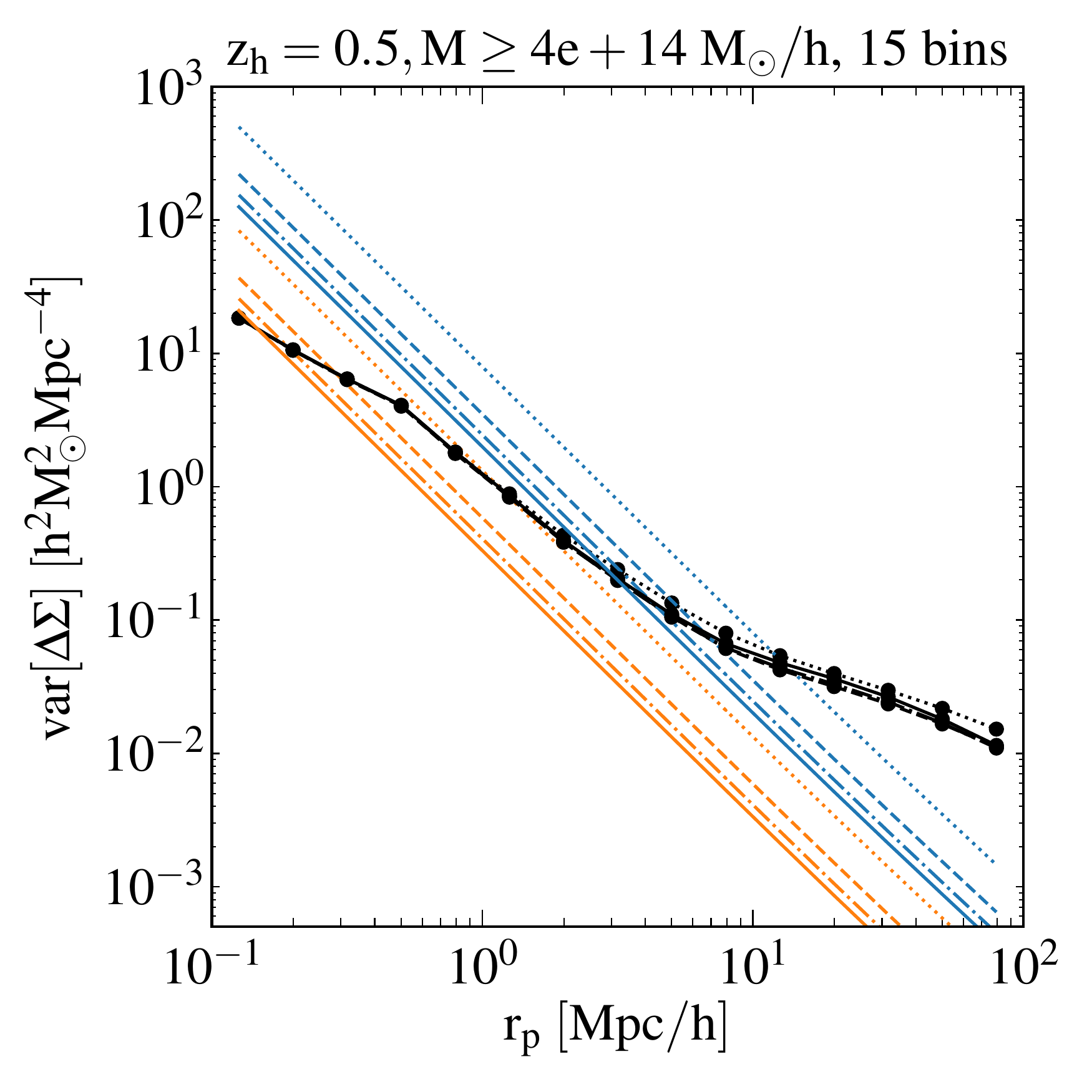}
\caption{Importance of shape noise compared with the shape noise-free calculation (the intrinsic halo profile variation at small radii and LSS at large radii).  We show the shape noise components for source densities 10 and 60 arcmin$^{-2}$ at various source redshifts. 
{Left}: $M\ge10^{14}\ \hiMsun$. 
{Right}: $M\ge4\times10^{14}\ \hiMsun$.
In both cases, shape noise is subdominant at large scale.  The relative importance of shape noise depends on halo mass and source redshift; for higher mass (right-hand panel) and higher source redshift (solid curves), shape noise is mostly subdominant for a source density 60 arcmin$^{-2}$.}
\label{fig:shape_noise_demo}
\end{figure*}
\begin{figure*}[H]
\includegraphics[width=\textwidth]{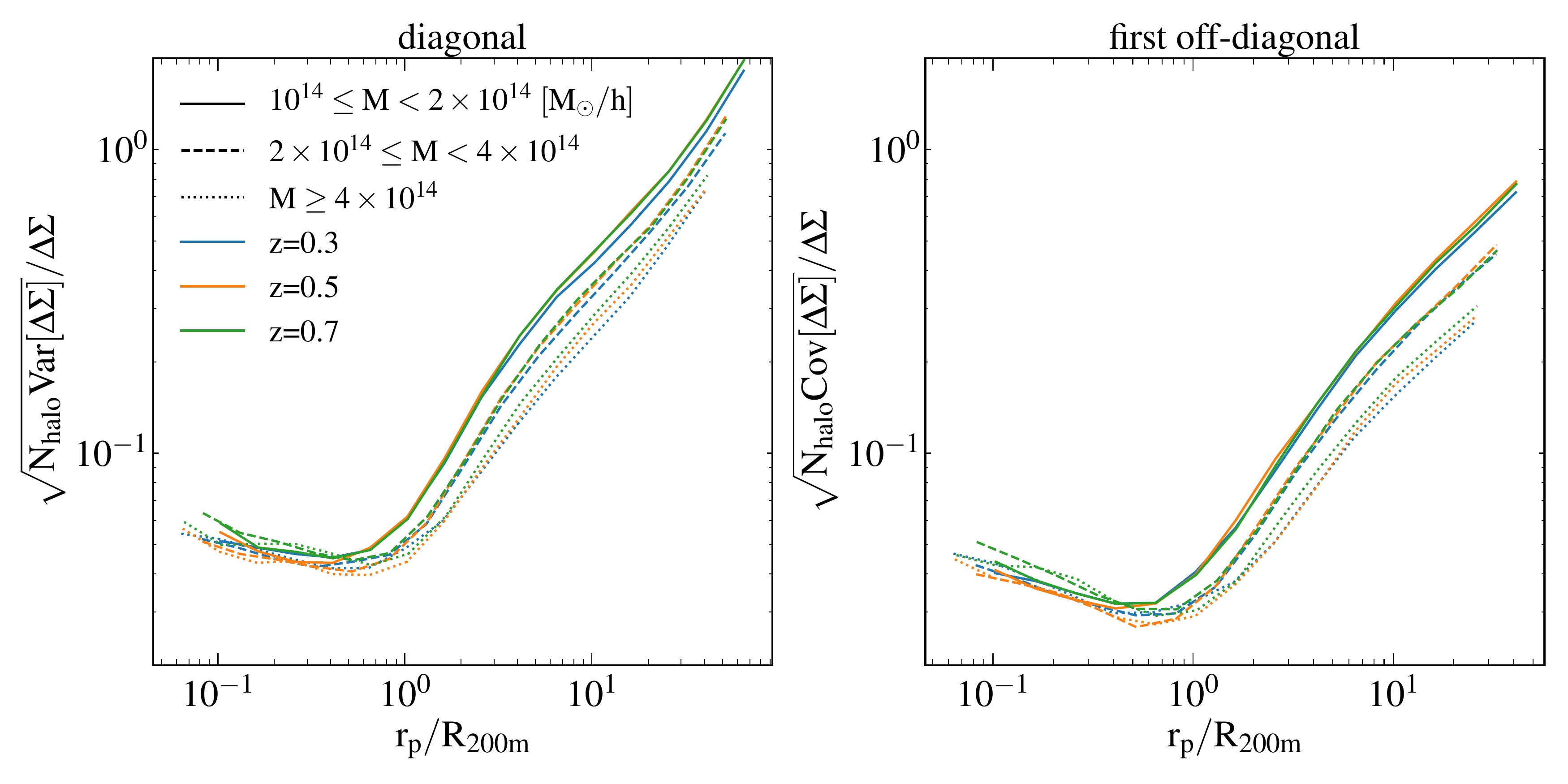}
\caption{Fractional errors of $\DS$ for various halo mass ranges and redshifts.  The left-hand panel shows the diagonal elements, while the right-hand panel shows the elements next to the diagonal.  Various colours indicate various redshifts, while various line styles indicate various mass ranges.  For a given mass range, the fractional errors of $\DS$ are approximately independent of redshift.  High mass haloes have slightly lower fractional errors (higher signal-to-noise ratio).}
\label{fig:fractional_error}
\end{figure*}
\begin{figure*}[H]
\includegraphics[width=\textwidth]{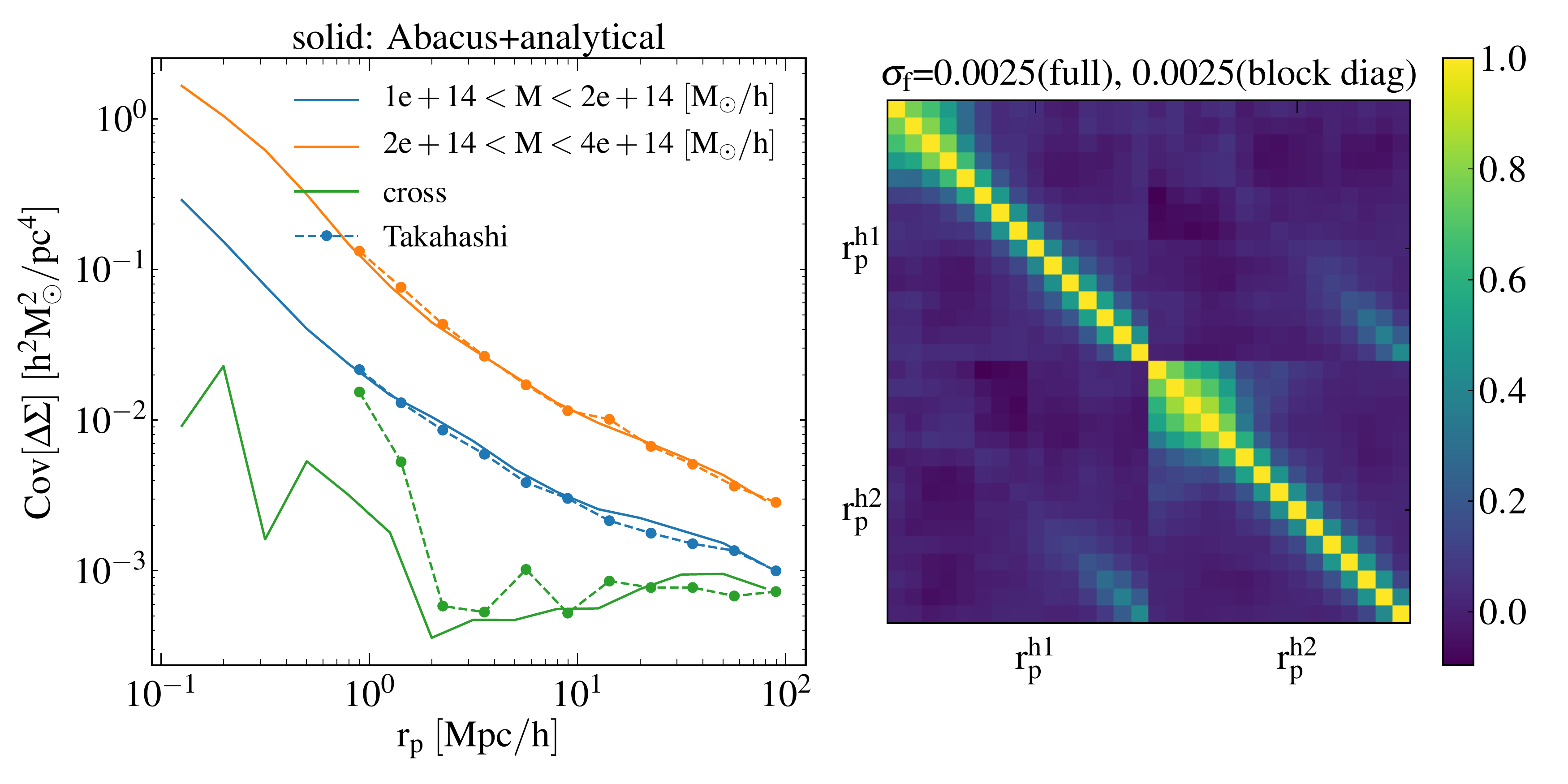}
\caption{
Cross-mass covariance matrix for $\DS$ for two mass bins: $[10^{14}, 2\times10^{14})$ and $[2\times10^{14}, 4\times10^{14})$ at $z=0.5$, in the case of negligible shape noise.
{Left}: diagonal elements of the covariance of each block.  
The blue curves correspond to the low-mass bin, the orange curves correspond to the high-mass bin, and the green curves correspond to the cross-mass covariance.
The solid curves are from Abacus+analytical, while the dashed curves are from Takahashi simulations.  At large radii, the cross-mass covariance can be modelled by LSS and the halo bias of the two mass bins.  At small radii, the cross-mass variance is approximately an order of magnitude lower than the variance of the low-mass bin.
{Right}: correlation matrix. The upper left block corresponds to the low-mass bin, the lower right block corresponds to the high-mass bin, and the upper right and lower left blocks correspond to the covariance between the two mass bins.}
\label{fig:cov_cross_mass}
\end{figure*}

Fig.~\ref{fig:shape_noise_demo} shows the relative importance of shape noise (various colour curves) and the shape noise-free part (black curves, we call it density variance hereafter) for the variance of $\DS$.  We use lenses at $z=0.5$ and 15 $\rp$ bins between 0.1 and 100 $\hiMpc$, and the two panels correspond to mass thresholds of $10^{14}$ and $4\times10^{14}\ \hiMsun$.  For lenses at $z=0.3$ and $z=0.7$, the results are nearly identical to the $z=0.5$ case shown here.   Various line styles correspond to various source redshifts.  For simplicity, we assume that all source galaxies are at the same redshift; in reality, the source galaxies are distributed in a redshift range.

For the density variance (black), the LSS contribution at large scales depends on the source redshift because we integrate all the line-of-sight structure in front of the source redshift. This redshift dependence is weak but non-monotonic.  The LSS contribution involves the ratio $\Sigcrit(\zs,\zh)/\Sigcrit(\zs,\zLSS)$; see equation~(\ref{eq:F_Sigma}).  As $\zs$ increases, both the numerator and the denominator decrease.  When $\zs$ is low and close to $\zh$ (e.g.~0.75), as $\zs$ increases the numerator decreases rapidly, leading to a smaller covariance.  When $\zs$ is high, as $\zs$ increases, the numerator is almost constant; since we integrate over a wider range of line-of-sight structure, the total covariance increases.

On the other hand, at a fixed source density, the shape noise is sensitive to the source redshift.  At a fixed $\zh$, $\Sigcrit$ decreases with $\zs$, and a higher $\zs$ corresponds to a lower shape noise due to the $\Sigcrit^2$ factor in equation~(\ref{eq:cov_DS}).  If we plot the variance of $\gamma_t$ instead, the density variance part would depend on the source redshift (because $\gamma_t$ is higher for a higher source redshift), while the shape noise would be independent of the source redshift (equation~\ref{eq:cov_gammat}).  In both cases, the relative importance of shape noise and density variance is the same.

The blue and orange curves correspond to source densities of 10 arcmin$^{-2}$ (DES-like) and 60 arcmin$^{-2}$ ({\rm WFIRST}-like).  For 10 arcmin$^{-2}$, the small-scale variance is dominated by shape noise.  For 60 arcmin$^{-2}$, small-scale density variance becomes more important.  Comparing the two panels, we can see that the density variance is more important for high-mass haloes.

When halo clustering is negligible, the shape noise is inversely proportional to the radial bin area.  For example, if we use 30 radial bins instead of 15, the shape noise would double, while the density variance would remain approximately the same.  With 30 radial bins, although the diagonal elements have larger shape noise, the off-diagonal elements play more important roles than the 15-bin case.

\subsection{Mass and redshift dependence}\label{sec:M_z_dependence}

In this section, we show that the fractional errors of $\DS$, in the case of negligible shape noise, are almost independent of redshift and only weakly depend on mass.  Fig.~\ref{fig:fractional_error} compares the fractional errors for $\DS$ from Abacus+analytical, 
for halo redshifts $\zh=$ 0.3, 0.5, and 0.7, 
and mass ranges 
$[10^{14},2\times10^{14})$, 
$[2\times10^{14},4\times10^{14})$, 
and $[4\times10^{14},\infty)$ $\hiMsun$.  
We assume source redshifts $\zs = 2.5 \zh$.
We multiply the covariance matrix of $\DS$ by the number of haloes in each bin. For the $y$-axis, we take the square root of this product and divide it by the mean of $\DS$.  For the $x$-axis, we divide the projected distance by the mean $R_{\rm 200m}$ in each bin.  The left-hand panel corresponds to the diagonal elements, and the right-hand panel corresponds to the elements next to the diagonal.

We use different colours to indicate different redshift bins, and different line styles to indicate different mass bins.  For a given mass bin (same line style), the fractional error profiles are almost independent of redshift.  Comparing different mass bins, we see that higher mass haloes have slightly lower fractional errors (higher signal-to-noise ratio). Both panels show very similar trends.   Our numerical calculations focus on specific mass and redshift bins, and the weak dependence demonstrated in Fig.~\ref{fig:fractional_error} can be useful for scaling our results to other redshift or mass bin choices, after which shape noise can be added using our analytical formulae.

\subsection{Cross-mass covariance}\label{sec:cov_cross_mass}

If we use multiple mass bins to perform stacked weak lensing measurements and to jointly constrain cosmological parameters, we need to take into account the covariance between mass bins.  To calculate the cross-mass covariance, we again combine Abacus simulations at small scales and analytical calculations at large scales.

To calculate the Gaussian-field covariance analytically for two mass bins at the same redshift (denoted as h1 and h2), we can write the covariance matrix analogous to equation (\ref{eq:cov_gammat}):
\beqa
&\Cov^{\rm Gauss}\bigg[ \gamma_t^{\rm h1}(\theta_1), \gamma_t^{\rm h2}(\theta_2) \bigg] 
= \frac{1}{4\uppi\fsky} \int \frac{\ell \dd\ell}{2\uppi}
\hJ_2(\ell \theta_1)\hJ_2(\ell \theta_2) \times
\\
&\left[
C_\ell^{\rm h1h2}\left(\clkk+\frac{\sigma_\gamma^2}{\ns}\right)    +
C_\ell^{h1 \kappa} C_\ell^{h2 \kappa}
\right]
\ . 
\label{eq:cov_cross_mass}
\eeqa
The cross-mass covariance has no shot noise associated with halo number counts, and the halo clustering is described by the cross-mass power spectrum
\beq
C_\ell^{\rm h1h2} = \int_{\chimin}^{\chimax} \dd\chih \left(\frac{F_{\rm h}(\chih)}{\chih} \right)^2
P_{\rm h1h2}\left(k=\frac{\ell{+1/2}}{\chih}\right) \ .
\label{eq:clh1h2}
\eeq
In the expression above, the 3D cross power spectrum of halo samples is given by
\beq
P_{\rm h1h2}(k) = b_1 b_2 P_{\rm lin}(k) \ ,
\eeq
where $b_1$ and $b_2$ are the halo bias values of the two samples.  Since we consider two samples at the same redshift, they have the same window function $F_h(\chih)$, which is given by equation~(\ref{eq:F_h}).  In this calculation, the large-scale cross-mass covariance matrix is symmetric.  The last term ($C_\ell^{h1 \kappa} C_\ell^{h2 \kappa}$) and the non-Gaussian part is calculated using Abacus simulations.

Fig.~\ref{fig:cov_cross_mass} shows the cross-mass covariance matrices for two mass bins: $[10^{14}, 2\times10^{14})$ and $[2\times10^{14}, 4\times10^{14})$ $\hiMsun$, for Abacus+analytical and Takahashi.  The left-hand panel shows the diagonal elements of each block.  The blue, orange, and green curves correspond to the low-mass bin, the high-mass bin, and the cross-mass covariance.  The solid curves correspond to the Abacus+analytical results, while the dash curves correspond to the Takahashi results.  The high-mass bin has a higher variance than the low-mass bin due to its higher shot noise.  At large scale, the cross-mass covariance is closer to the low-mass bin because both are dominated by halo clustering.  For the cross-mass variance, shot noise has no contribution; for the low-mass bin, halo clustering is higher than shot noise; for the high-mass bin, shot noise dominates.  At small scales, the cross-mass covariance is approximately an order of magnitude smaller than the covariance of the low-mass bin.  

The right-hand panel shows the full correlation matrix. The upper left block corresponds to the low-mass bin (15 radial bins between 0.1 and 100 $\hiMpc$), and the lower right block corresponds to the high-mass bin calculated with the same radial binning. The upper right and lower left blocks correspond to the cross-mass covariance of the two mass bins and are the transpose of each other.

To understand the contribution of the cross-mass block to the error budget, we again calculate the constraints on a parameter multiplying the $\DS$ for both mass bins.  We combine the two data vectors to form 
\[
  \DS=
  \left[ {\begin{array}{c}
   \DS(\mbox{bin 1})\\
   \DS(\mbox{bin 2})\\
  \end{array} } \right]  \ ,
\]
and we again use equation~(\ref{eq:sigmaf}).  We calculate $\sigmaf$ using the full matrix vs.\ a block-diagonal matrix (ignoring the cross-mass covariance).  Both calculations give $\sigmaf = 0.0025$ for a 1 $(\hiGpc)^3$ volume, indicating that the cross-mass covariance is negligible in the absence of shape noise.  However, if shape noise dominates, we need to add diagonal shape noise to the cross-mass covariance block.

\subsection{Potential systematic uncertainties}\label{sec:systematics}

In this section we discuss the impact of baryons, cluster miscentering, and mass-observable relation on the covariance calculations.

In this work, the small-scale cluster lensing signal is derived from dark matter-only simulations. Baryonic effects can change the inner density profiles of clusters \citep[e.g.][]{Duffy10, Martizzi12, Schaller15}  and thus change the lensing signal.  Since the effect is mainly at very small scales (less than 10 kpc), which will be dominated by shape noise for most of the surveys, its impact is likely to be small for covariance matrices.

We calculate the cluster lensing signals with respect to the centres of dark matter haloes. In optically-selected cluster samples, the locations of the brightest cluster galaxies may not coincide with the centres of dark matter haloes (see e.g.~\citealt{Zhang19} for comparisons between optically defined centres and X-ray centres).  This miscentering would lead to shallower lensing profiles at small scales.  For current surveys, the scales affected by miscentering are dominated by shape noise, and the covariance matrices are unaffected.  For future surveys, when shape noise no longer dominates at small scales, the miscentering effects would need to be taken into account in modelling the small-scale covariance matrices.

In this work we select clusters based on their masses; in real surveys, clusters would be selected by some observed property (e.g.~optical richness), which has a scaling relation and a scatter around the true mass.  If this mass-observable relation is not biased by some property that can also bias lensing, we can simply incorporate this effect by convolving the halo mass function with the mass-observable relation.  However, if the mass-observable relation is biased by some property that can also bias the lensing signal, for example, the orientation of a triaxial halo, we will need to take into account this extra property in order to obtain unbiased lensing measurements and robust covariance matrices.  We will explore this in future work.

\section{Summary}\label{sec:summary}

We calculate accurate covariance matrices required for the cosmological interpretation of weak gravitational lensing by galaxy clusters that will be observed by LSST, {\rm Euclid}, and {\rm WFIRST}.  We combine analytical calculations at large scales with the Abacus N-body simulations at small scales, and we validate our approach with the Takahashi full-sky ray-tracing simulations at intermediate and large scales.  Our main results are summarised as follows.

\bit

\item We demonstrate two pitfalls of calculating and interpreting cluster lensing covariance matrices: (1) the importance of subtracting random lensing signals, and (2) the difference between the halo-to-halo covariance and the patch-to-patch covariance. The latter is relevant for stacked cluster lensing analyses.

\item We analytically calculate the contribution from shape noise and uncorrelated LSS to lensing covariance assuming Gaussian random fields.  This part uses the angular power spectra of haloes and lensing.  The uncorrelated LSS dominates the large-scale covariance.

\item For the small-scale covariance, we use the Abacus Cosmos N-body simulations to calculate the contribution from the intrinsic variation of halo density profiles.   We combine this small-scale covariance with the large-scale Gaussian-field calculation and validate these calculations using the simulated full-sky lensing maps by Takahashi.

\item  In the absence of shape noise, the diagonal elements of the covariance matrix are insensitive to the radial bin size.  The off-diagonal elements associated with two scales are also insensitive to the radial bin size. Narrowing radial bin width increases the number of correlated bins while conserving the constraints on lensing amplitudes.

\item In the absence of shape noise, ignoring the off-diagonal elements can lead to approximately a factor of two underestimation on the constraints on lensing amplitudes.  Including off-diagonal elements that are separated by approximately 0.6 dex in radius would avoid this underestimation.  Off-diagonal elements separated by more than 0.6 dex in radius have negligible impact on these constraints.

\item We compare the relative importance between shape noise and density variance.  The details depend on the source redshift, radial bin, and halo mass.  In general, for a source density of 10 arcmin$^{-2}$, shape noise is subdominant for $\rp \ga 5\ \hiMpc$; for 60 arcmin$^{-2}$, shape noise becomes subdominant at most scales.

\item In the absence of shape noise, the fractional error of cluster lensing is independent of cluster redshift and only weakly depends on mass.  Higher mass clusters have slightly higher signal-to-noise ratio.

\item We investigate the cross-covariance of two halo samples of different mass ranges in the same redshift, same survey region.  When halo shot noise dominates ($1/\nh \gg \clhh$) and when shape noise is low, the cross-mass covariance blocks are significantly smaller than the auto-mass covariance blocks and can be safely ignored when we constrain parameters. 

\eit 

The covariance matrices provided in this paper can be used for robust forecast and survey optimisation for future surveys like LSST, {\rm Euclid}, and {\rm WFIRST}.  Given the high signal-to-noise ratio of these surveys, cluster lensing will have unprecedented constraining power on the growth of structure. In Wu et al.\ (in preparation) and \cite{Salcedo19}, we present forecasts on combining cluster counts and lensing, as well as how we can effectively combine the cluster lensing and cross correlation functions of clusters and galaxies to constrain cosmological parameters. 

\section*{Acknowledgements}

We gratefully acknowledge use of the Abacus Cosmos simulation suite and thank Lehman Garrison and Daniel Eisenstein for advice on its application.
We thank Ryuichi Takahashi and his team for providing the ray-tracing simulations, and 
Elisabeth Krause and Tim Eifler for providing the {\sc Cosmolike} software.
This work was supported in part by NSF Grant AST-1516997 and NASA Grant 15-WFIRST15-0008.
ANS is supported by the Department of Energy Computational Science Graduate Fellowship Program under contract DE-FG02-97ER25308.
BDW is supported by the National Science Foundation Graduate Research Fellowship Program under Grant No.\ DGE-1343012.  
The computations in this paper were run on the CCAPP condo of the Ruby Cluster at the Ohio Supercomputer Center.

\bibliographystyle{mnras}
\bibliography{master_refs}

\begingroup
\let\oldclearpage\clearpage
\let\clearpage\relax 
\onecolumn


\appendix

\section{Derivation of the Gaussian-field covariance matrices}\label{app:cov}

The mean of $\gamma_t$ can be written as the Hankel transform of  $\clhk$ (the cross angular power spectrum of halo and convergence):
\beq
\avg{\gamma_t}(\theta) = \int\frac{\ell \dd\ell}{2\uppi} \clhk J_2(\ell \theta) \ .
\eeq
To calculate the covariance of $\gamma_t$, let us first focus on the covariance of $C_\ell^{h\kappa}$, which is given by \citep[see e.g.][]{Hu04,Krause17}:
\beq
\Cov\bigg[C_{\ell}^{h\kappa}, C_{\ell'}^{h\kappa}\bigg] = 
\frac{\delta_{\ell\ell'}}{ \fsky (2\ell+1)\Delta\ell }
\bigg[
  (C_{\ell}^{hh}+ N_h)(C_{\ell}^{\kappa\kappa}+ N_\kappa) +(C_{\ell}^{h\kappa})^2 
\bigg] \ .
\eeq
We perform Hankel transform in the form of Riemann sum twice to obtain the covariance of $\gamma_t$ in real space:
\beqa
\Cov\bigg[ \gamma_t(\theta), \gamma_t(\theta') \bigg] 
&=\frac{1}{(2\uppi)^2} \sum_\ell \ell \Delta\ell \sum_{\ell'}\ell' \Delta\ell' 
J_2(\ell\theta)J_2(\ell'\theta')  \Cov\bigg[C_{\ell}^{h\kappa}, C_{\ell'}^{h\kappa}\bigg] \\
&  = \frac{1}{8\uppi^2\fsky}\sum_\ell \ell \Delta\ell J_2(\ell\theta)J_2(\ell\theta')
  \bigg[
   (C_{\ell}^{hh}+ N_h)(C_{\ell}^{\kappa\kappa}+ N_\kappa) +(C_{\ell}^{h\kappa})^2
 \bigg] \\
&= \frac{1}{4\uppi\fsky} \int \frac{\ell \dd\ell}{2\uppi}  J_2(\ell \theta)J_2(\ell \theta')
\bigg[   
\left(C_{\ell}^{hh} + \frac{1}{\nh}\right)
\left(C_{\ell}^{\kappa\kappa} + \frac{\sigma_\gamma^2}{\ns}\right) +(C_{\ell}^{h\kappa})^2 
\bigg]  \ .
\eeqa 
In the derivation above we assume $\ell \gg 1$ and replace the Riemann sum with an integral; $\nh$ and $\ns$ are the surface number densities of haloes and sources in the unit of sr$^{-1}$.

\section{Radial bin-averaged Bessel function}\label{app:bessel}

\begin{figure}
\centering
\includegraphics[width=0.5\textwidth]{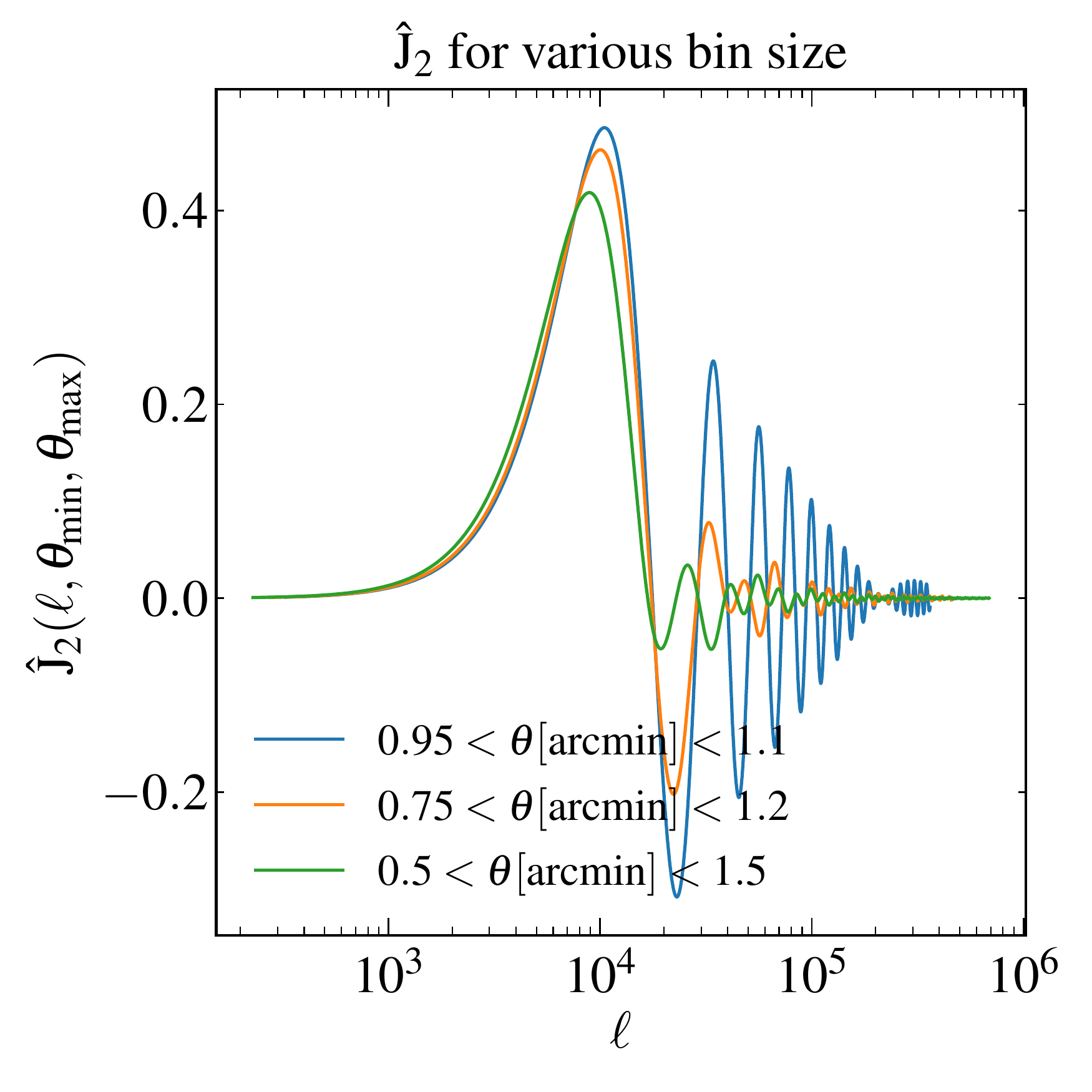}
\caption{Radial bin averaged Bessel function $\hJ_2$ (equation~\ref{eq:hJ2}) for various bin widths with a fixed bin centre; $J_2$ peaks at $\ell \theta$ = 2, and the first peak only weakly depends on the bin size. For a larger bin size, the function decays more rapidly.}
\label{fig:bessel_demo}
\end{figure}

When we calculate the real-space covariance by integrating the angular power spectra, the size of the radial bin is taken into account by the bin-averaged Bessel function.  
For a finite radial bin, equation~(\ref{eq:cov_gammat}) uses the average of $J_2$ within the radial bin $\thetamin < \theta < \thetamax$,
\beqa
\ \hJ_2(\ell,\thetamin, \thetamax)
=& \frac{1}{\uppi(\thetamax^2 - \thetamin^2)} \int_{\thetamin}^{\thetamax} J_2(\ell \theta) 2\uppi \theta \dd\theta   \\
=&  \frac{2}{ \ell^2 (\thetamax^2 - \thetamin^2)}   
 \bigg[
 2 \bigg( J_0( \ell \thetamin)-J_0(\ell \thetamax)\bigg)
 + \ell \bigg( \thetamin J_1(\ell \thetamin) -\thetamax J_1(\ell \thetamax) \bigg)
 \bigg] \ .
\label{eq:hJ2}
\eeqa
One can obtain the last expression using the recurrence relation of Bessel function or using {\sc Mathematica}.  Fig.~\ref{fig:bessel_demo} shows examples of $\hJ_2$ at a fixed bin centre with different bin widths. The first peak, which occurs at $\ell\theta = 2$, is almost independent of the bin width.  For a larger bin width, $\hJ_2$ decays more rapidly with $\ell$.  If $\hJ_2$ is convolved with a function that decreases rapidly with $\ell$, only the first peak of $\hJ_2$ would contribute significantly, and the results would be almost independent of the bin size (as is the case for LSS).  On the other hand, if $\hJ_2$ is convolved with a function with weak or no scale dependence, the results would be larger for a smaller bin size (as is the case for shot noise and shape noise).

Let us denote the area of the radial bin as
\beq
A_{\rm ann} = \uppi(\thetamax^2 - \thetamin^2) \ .
\eeq
When the halo shot noise dominates ($1/\nh \gg \clhh$), 
the diagonal elements of the shape noise (equation~\ref{eq:cov_shape}) only involve this integral
\beq
\int \frac{\ell \dd\ell}{2\uppi} 
\left[ \hJ_2(\ell, \thetamin, \thetamax) \right] ^2  
=\frac{1}{A_{\rm ann}^2}
\int_{\thetamin}^{\thetamax} 2\uppi \theta \dd\theta
\int_{\thetamin}^{\thetamax}  2\uppi \theta^{\prime} \dd\theta^{\prime} \int \frac{\ell \dd\ell}{2\uppi} J_2(\ell\theta) J_2(\ell\theta^{\prime}) 
= \frac{1}{A_{\rm ann}} \ ,
\eeq
which is exactly one over the area of the radial bin.  
In the derivation above we use the identity
\beq
\int \ell \dd\ell J_\alpha(\ell \theta) J_\alpha(\ell\theta') = \frac{\delta(\theta-\theta')}{\theta'}   \ .
\eeq
This expression would be zero if the integration limits of $\theta$ and $\theta'$ do not overlap (which corresponds to the off-diagonal elements).  Therefore, when halo clustering is negligible, shape noise only contributes to the diagonal elements of the covariance matrices.   On the other hand, when the Bessel integration has extra $\ell$-dependence in the integrand, the covariance matrices would become non-diagonal, as is the case for non-negligible halo clustering and LSS.

\section{Angular power spectra from the Limber approximation}\label{app:limber}
\begin{figure}
\centering
\includegraphics[width=0.5\textwidth]{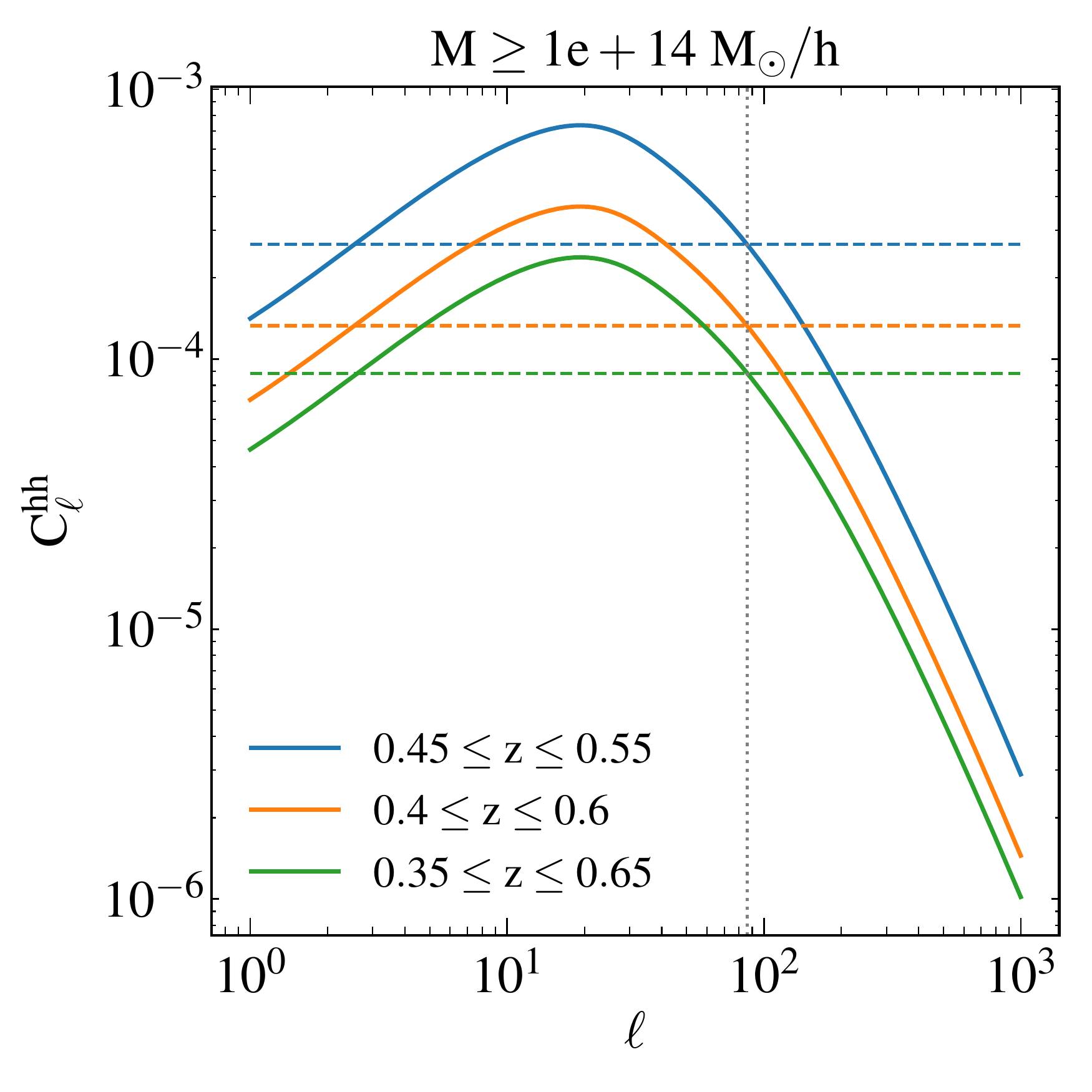}
\caption{Halo clustering angular power spectra (solid curves) and shot noise (horizontal dash lines) for various redshift bin sizes.  A larger redshift bin corresponds to a lower angular power spectrum and lower shot noise, while the transition scale at which shot noise starts to dominate (vertical line) remains constant.}
\label{fig:demo_shot_noise_vs_clustering_clhh_redshift_binsize}
\end{figure}

We use the Limber approximation to compute the angular power spectrum of a 3D density field, assuming flat sky and small angular separation. 
We follow the treatment in \cite{LoverdeAfshordi08} and summarise the key equations below. 
If a 2D field $A^{\rm 2D}$ is related to the underlying 3D field $A^{\rm 3D}$ via a dimensionless window function $F_A$,
\beq
A^{\rm 2D}(\vec{n}) = \int \dd\chi F_{\rm A}(\chi) A^{\rm 3D}(\vec{x})  \ ,
\eeq
then the angular power spectrum is related to the 3D power spectrum via
\beq
C_\ell^{\rm A} = \int \dd\chi  \left(\frac{F_{\rm A}(\chi)}{\chi} \right)^2 P_{\rm A}\left(k=\frac{\ell{+1/2}}{\chi}\right) \ .
\eeq
This expression can be generalised to the cross power spectrum between two quantities A and B
\beq
C_\ell^{\rm AB} = \int \dd\chi  \left(\frac{F_{\rm A}(\chi)}{\chi} \right) \left(\frac{F_{\rm B}(\chi)}{\chi} \right) P_{\rm AB}\left(k=\frac{\ell{+1/2}}{\chi}\right)  \ .
\eeq

For the halo sample in a redshift slice corresponding to the comoving distance range ($\chimin$, $\chimax$), the 2D number density is given by
\beq
\delta_h^{\rm 2D} = \int_{\chimin}^{\chimax} \dd\chih \delta_h^{\rm 3D} \frac{\chih^2}{V}  \ ,
\eeq
where
\beq
V = \int_{\chimin}^{\chimax} \dd\chih \chih^2 \ ;
\eeq
therefore, the window function is given by
\beq
F_{\rm h}(\chih) = \frac{\chih^2}{V} \ ,
\eeq
and the power spectrum is given by
\beq
\clhh = \int_{\chimin}^{\chimax} \dd\chih \left(\frac{F_{\rm h}(\chih)}{\chih} \right)^2 P_{\rm hh}\left(k=\frac{\ell{+1/2}}{\chih}\right) \ .
\eeq

Fig.~\ref{fig:demo_shot_noise_vs_clustering_clhh_redshift_binsize} shows the effect of redshift bin size on $\clhh$ and $1/\nh$; namely, the relative importance of halo clustering and shot noise as a function of scale.  We use a halo mass threshold of $10^{14}\ \hiMsun$.  When we use a wider redshift bin, both terms decrease, but the transition $\ell$ scale where shot noise starts to dominate remains the same.

For the convergence, we first integrate the LSS along the line-of-sight in front of a given source redshift and then integrate the source distribution $\psrc(\chis)$.  We then swap the order of integrations of $\chis$ and $\chiLSS$ and obtain
(see e.g.~equation 21 in \citealt{Kilbinger15}):
\beq
\kappa = \int_0^\infty \dd\chiLSS \delta_m^{\rm 3D} 
\bar{\rho}
\int_{\chiLSS}^{\infty} \dd\chis \psrc(\chis)
\frac{1}{\Sigcrit(\zs, \zLSS)}   \ ;
\eeq
therefore, the window function is given by
\beq
F_\kappa(\chiLSS) = \bar{\rho} \int_{\chiLSS}^{\infty} 
\dd\chis \psrc(\chis)\frac{1}{\Sigcrit(\zs, \zLSS)} \ ,
\eeq
and the angular power spectrum is given by
\beq
\clkk = \int_0^\infty \dd\chiLSS \left(\frac{F_{\kappa}(\chiLSS)}{\chiLSS} \right)^2 P_{\rm mm}\left(k=\frac{\ell{+1/2}}{\chiLSS}\right) \ . 
\eeq
The cross angular power spectra between halo and $\kappa$ is given by
\beq
\clhk = \int_{\chimin}^{\chimax} \dd\chi \left(\frac{F_{\rm h}(\chi)}{\chi} \right)
\left(\frac{F_{\rm \kappa}(\chi)}{\chi} \right)
P_{\rm hm}\left(k=\frac{\ell{+1/2}}{\chi}\right) \ ,
\eeq
where we integrate the matter distribution over the haloes' redshift range.

For $\Sigma$, it is equivalent to multiplying the $\kappa$ of each source by $\Sigcrit(\zs, \zh)$, where $\zh$ is the redshift of haloes:
\beq
\Sigma(\zh) = \int_0^\infty \dd\chiLSS \delta_m^{\rm 3D} 
\bar{\rho}
\int_{\chiLSS}^{\infty} \dd\chis \psrc(\chis) \frac{\Sigcrit(\zs, \zh) }{\Sigcrit(\zs, \zLSS)} \ .
\eeq
This expression only applies to a narrow halo redshift bin.
The window function is thus given by
\beq
F_\Sigma(\chiLSS, \zh) = \bar{\rho} \int_{\chiLSS}^{\infty} \dd\chis 
\psrc(\chis)
\frac{\Sigcrit(\zs, \zh)}{\Sigcrit(\zs, \zLSS)}  \ .
\eeq
This window function is not dimensionless because of the extra $\Sigcrit$.
The angular power spectrum is thus given by
\beq
\clSS(\zh) = \int_0^\infty \dd\chiLSS 
\left(\frac{F_{\Sigma}(\chiLSS, \zh)}{\chiLSS} \right)^2
P_{\rm mm}\left(k=\frac{\ell{+1/2}}{\chiLSS}\right)  \ .
\eeq
The cross power spectrum between haloes and $\Sigma$ is given by
\beq
\clhS = \int_{\chimin}^{\chimax} \dd\chi 
\left(\frac{F_{\rm h}(\chi, \zh)}{\chi} \right) \left(\frac{F_{\Sigma}(\chi)}{\chi} \right)
P_{\rm hm}\left(k=\frac{\ell{+1/2}}{\chi}\right) \ .
\eeq

\section{Halo model used in this paper}\label{app:halo_model}

For calculating $\clhk$ (equation~\ref{eq:clhk}) and $\clhS$ (equation~\ref{eq:clhS}), we need the halo-matter power spectrum calculated from the halo model \citep[e.g.][]{CooraySheth02,Krause17}.  The total power spectrum includes the contribution from the 1-halo term and the 2-halo term 
\beqa
P_{\rm hm} (k)&=\Phm^{1h}(k)  +\Phm^{2h}(k) \\
P_{\rm hm}^{2h}(k) &= \bar{b} P_{\rm lin}(k) \\
P_{\rm hm}^{1h}(k) &= \frac{\int \dd M\frac{dn}{dM} \left(\frac{M}{\rho}\right) \tu(k) }{ \int \dd M\frac{dn}{dM}}  \ .
\eeqa
To calculate the 1-halo contribution,  we need the Fourier transform of the halo density profile, $\tu(k)$.  For an NFW profile, $\tu(k)$ has an analytical expression (e.g.~\citealt{CooraySheth02} equation 81)
\beq
\ \tu_m(k) =  \frac{4\uppi \rho_s r_s^3}{m}
\bigg\{ \sin(k r_s)\bigg\lbrack \Si\bigg((1+c)kr_s\bigg) 
- \Si(k r_s) \bigg\rbrack 
- \frac{\sin(ckr_s)}{(1+c)kr_s}
+\cos(k r_s) \bigg\lbrack \Ci\bigg((1+c)kr_s\bigg) - \Ci(k r_s) \bigg\rbrack
\bigg\}  \ ,
\eeq
where
\beqa
\Ci &= -\int_x^\infty \frac{\cos t}{t}\dd t\\
\Si &= \int_0^x \frac{\sin t}{t} \dd t  \ .
\eeqa

\let\clearpage\oldclearpage
\clearpage
\section{Four terms contributing to the Gaussian-field covariance}\label{app:demo_shot_noise_vs_clustering}
\begin{figure*}
\includegraphics[width=0.48\textwidth]{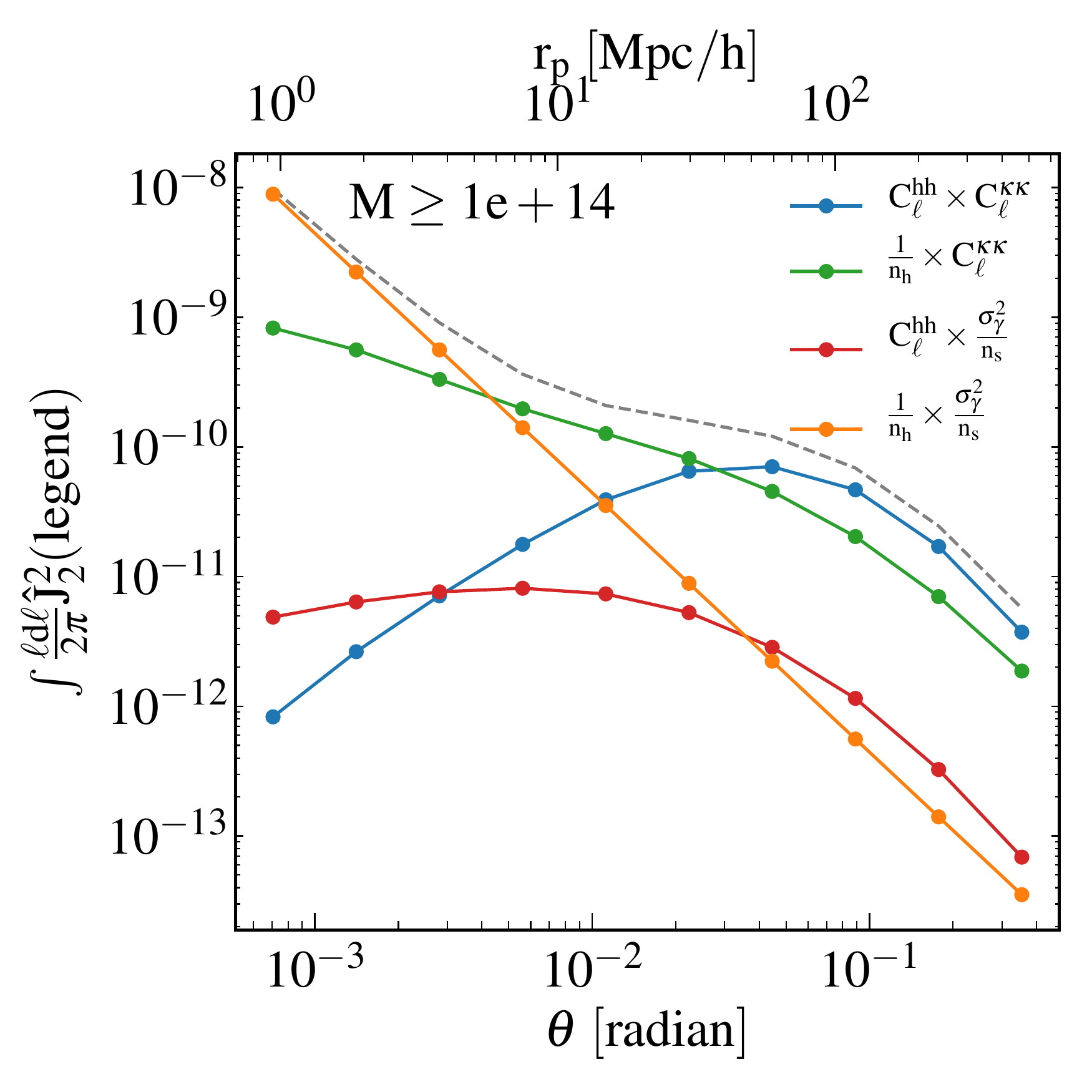}
\includegraphics[width=0.48\textwidth]{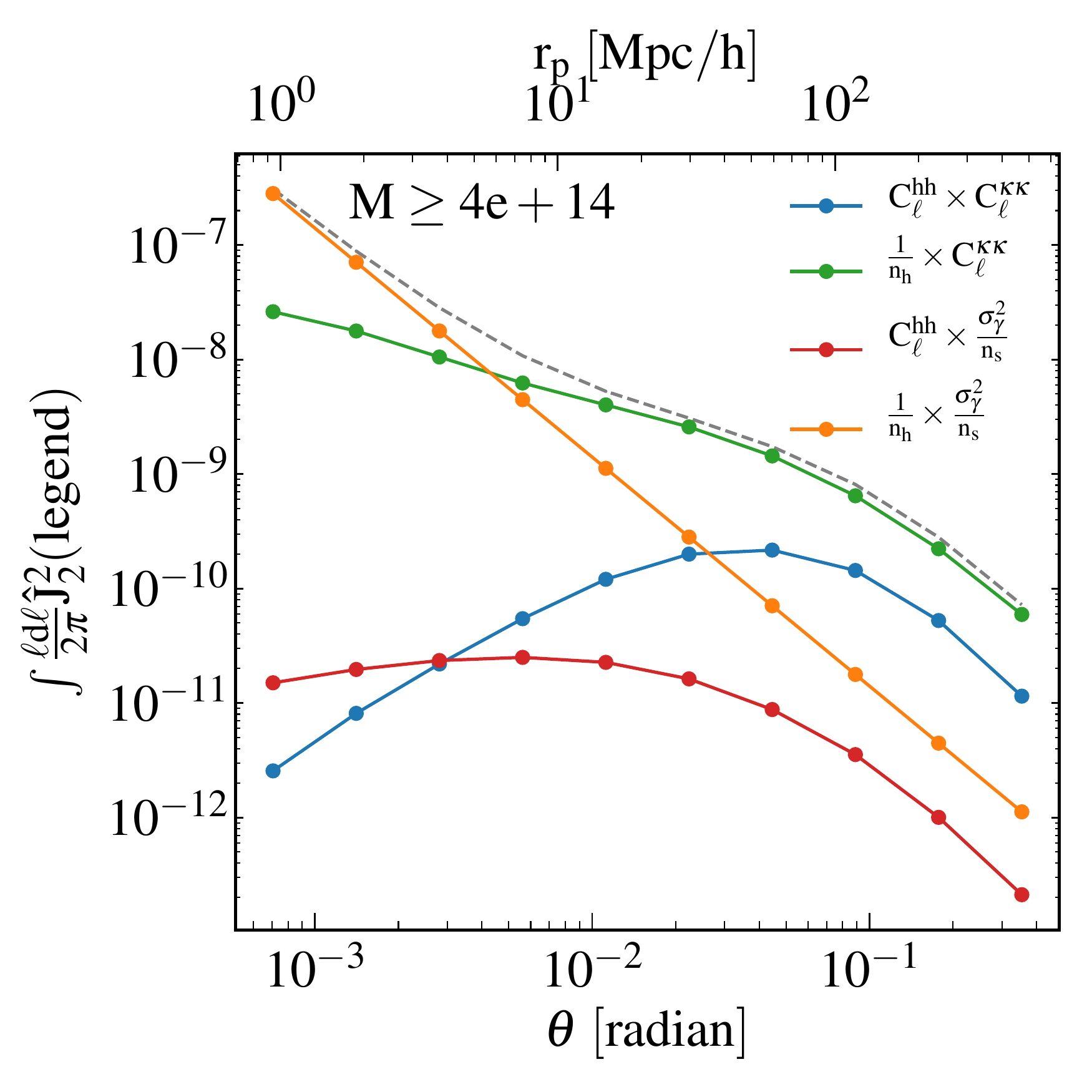}
\caption{Relative importance of the four terms contributing to the Gaussian-field covariance matrix, for two halo mass thresholds, at $z=0.5$.  We assume a source density 10 arcmin$^{-2}$ at $\zs=1.25$.  Left: $10^{14}\ \hiMsun$; right: $4\times10^{14}\ \hiMsun$.  The blue and green curves show the relative importance of halo clustering and shot noise; the former dominates at a lower mass threshold.  The green and orange curves show the relative importance between LSS and shape noise; the former dominates at large scales.}
\label{fig:demo_shot_noise_vs_clustering}
\end{figure*}
Fig.~\ref{fig:demo_shot_noise_vs_clustering} compares the four terms in the Bessel integration in equation~(\ref{eq:cov_gammat}):
\begin{enumerate}
\item  $\clhh \clkk$: halo clustering and LSS (blue)
\item  $\frac{1}{\nh} \clkk$: halo shot noise and LSS (green)
\item  $\clhh \frac{\sigma_\gamma^2}{\ns}$: halo clustering and source shape noise (red)
\item  $\frac{1}{\nh} \frac{\sigma_\gamma^2}{\ns}$: halo shot noise and source shape noise  (orange)
\end{enumerate}
We compare two mass thresholds: $10^{14}$ and $4\times10^{14}\ \hiMsun$.
We assume $\zh=0.5$, $\zs=1.25$, and a source density 10 arcmin$^{-2}$.
Terms (i) and (ii) correspond to the relative importance between halo clustering and shot noise.  The shot noise is higher than halo clustering for a higher mass threshold. At large scales,  the blue curve (Term i) is higher than the green curve (Term ii) for $M\ge10^{14}\ \hiMsun$ (left-hand panel) and is lower for  $M\ge4\times10^{14}\ \hiMsun$ (right-hand panel).  The relative importance between Term (iii) and Term (iv) follows the same reason.

Terms (ii) and (iv) correspond to the relative importance between LSS and shape noise.  At large radii, as shown in Fig.~\ref{fig:cl_takahashi}, the LSS contribution dominates, and therefore Term (ii) plotted in green is higher than Term (iv) plotted in orange.  However, at small radii, the shape noise dominates and is inversely proportional to the bin area, and therefore (iv) scales as a power-law and is higher than (ii).

\section{Unsuccessful attempt to calculate non-Gaussian covariance analytically}\label{app:non_gaussian}

\begin{figure}
\centering
\includegraphics[width=0.5\textwidth]{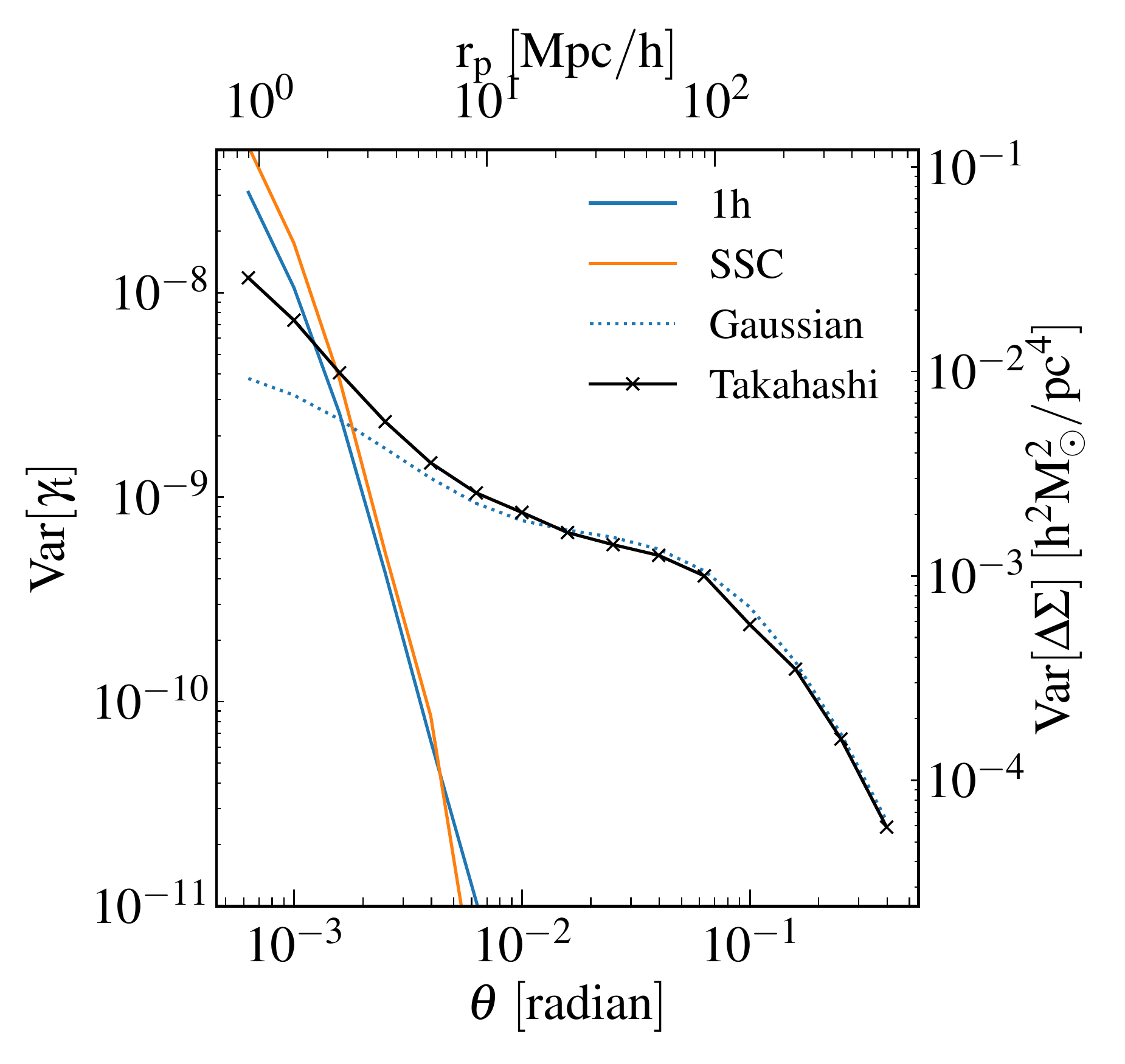}
\caption{Analytical calculation for small-scale non-Gaussian covariance including the 1-halo trispectrum (blue) and SSC (orange).  The non-Gaussian model overpredicts the small-scale variance in the Takahashi simulations (black).}
\label{fig:cov_ng}
\end{figure}

The analytical non-Gaussian covariance matrix includes two components: the connected four-point function (trispectrum) and the super sample covariance (SSC).

The contribution from the trispectrum is given by \citep[see e.g.][]{CoorayHu01,Takada09,Krause17}
\beq
\Cov^{\rm tri}\bigg[ \gamma_t(\theta_1), \gamma_t(\theta_2) \bigg] 
  =\frac{1}{4\uppi\fsky} \int \frac{\ell d\ell}{2\uppi} \int \frac{\ell d\ell}{2\uppi} \hJ_2(\ell\theta)\hJ_2(\ell'\theta)
T_{h\kappa h \kappa}(\ell, -\ell, \ell', -\ell')  \ ,
\eeq
where $T_{h\kappa h \kappa}$ is the angular trispectrum defined as
\beq
\avg{\delta_h(\ell)\kappa(-\ell)\delta_h(\ell')\kappa(-\ell')} = (2\uppi)^2
T_{h\kappa h \kappa}(\ell, -\ell, \ell', -\ell')  \ .
\eeq
We use the Limber approximation to calculate the angular trispectrum from the 3D trispectrum,
\beq
T_{h\kappa h \kappa}(\ell, -\ell, \ell', -\ell') 
= \int d\chi  \frac{F_h(\chi)^2 F_\kappa(\chi)^2}{\chi^6} T_{hmhm}(\ell/\chi, -\ell/\chi, \ell'/\chi, -\ell'/\chi)  \ .
\eeq
Similar to the case of power spectrum, we integrate over the redshift bin of the lenses. 

The 3D trispectrum at small scale is dominated by the 1-halo term, which is given by
\beq
T_{hmhm}(k, -k, k', -k') 
= \frac{1}{n_h^2}\int dM \frac{dn}{dM} \left(\frac{M}{\bar{\rho}}\right)^2 u(k) u(k')
\ ,
\eeq
where
\beq
n_h = \int dM\frac{dn}{dM} \ .
\eeq
Fig.~\ref{fig:cov_ng} shows that the 1-halo trispectrum overestimates the small-scale variance.  
In order to validate our implementation, we have performed detailed comparisons with the {\sc Cosmolike} software in $\ell$-space.  
To our knowledge, the trispectrum contribution has never been compared with cluster lensing simulations in the literature (but see \citealt{Takahashi19} for galaxy-galaxy lensing).
Our unsuccessful attempt indicates that the 1-halo trispectrum does not provide an adequate description for the small scale structures.  Readers with difference experiences are encouraged to contact the authors.

The SSC for cluster lensing is given by \citep[see e.g.][]{TakadaHu13,Krause17}
\beq
{\Cov}^{\rm SSC}(\ell_1, \ell_2)  = \int d\chi  \frac{F_h(\chi)^2 F_\kappa(\chi)^2}{\chi^4}
\left(\frac{\partial P_{h\kappa}(\ell_1/\chi)}{\partial \delta_b}\right)
\left(\frac{\partial P_{h\kappa}(\ell_2/\chi)}{\partial \delta_b}\right) 
\sigma_b(\Omega_s, z)  \ .
\eeq
Here $\sigma_b$ is the fluctuation in the survey window and depends on the survey area ($\Omega_s=4\uppi\fsky$); for a circular survey area, $\sigma_b$ is given by
\beq 
\sigma_b(\Omega_s, z) = \int \frac{d^2{\bf k_\perp}}{(2\uppi)^2} P_{\rm lin}(k_\perp,z)
\left[\frac{2 J_1(k_\perp \chi \theta_s)}{k_\perp \chi \theta_s} 
\right]^2
\ ,
\eeq
and $\theta_s = \sqrt{\Omega_s/\uppi}$
\citep[see e.g.][for detailed implementations]{Krause17}.
We use the implementation in {\sc Cosmolike} for a sky coverage of $\fsky = 1/48$.
We then perform the Hankel transform twice to calculate the real-space covariance,
\beq
\Cov^{\rm ssc}\bigg[ \gamma_t(\theta_1), \gamma_t(\theta_2) \bigg] 
=  \int \frac{\ell_1 d\ell_1}{2\uppi}
\int \frac{\ell_2 d\ell_2}{2\uppi}
{\Cov}^{\rm SSC}(\ell_1, \ell_2) \hJ_2(\ell_1\theta_1) \hJ_2(\ell_2\theta_2) \ .
\eeq
Fig.~\ref{fig:cov_ng} shows the resulting SSC contribution to the variance, which is higher than the simulation results. The discrepancy could be related to the lack of SSC in the simulation.  Nevertheless, the discrepancy between the analytical non-Gaussian model and simulations does not affect the results in this paper.

\endgroup

\bsp	
\label{lastpage}
\end{document}